\documentclass[superscriptaddress,amsmath,amssymb,twocolumn]{revtex4-2}

\usepackage{graphicx}%
\usepackage{dcolumn}%
\usepackage{bm}%
\usepackage{bbm}
\usepackage{braket}
\usepackage{amsmath}
\usepackage{empheq}
\usepackage{mathtools}
\usepackage{xparse}
\usepackage{xcolor}
\usepackage[normalem]{ulem}
\usepackage{mathrsfs}

\makeatletter
\renewcommand\frontmatter@makefntext[1]{%
  \noindent\@makefnmark\ #1%
}
\makeatother

\let\latexchi\chi
\makeatletter
\DeclareRobustCommand\chi{\@ifnextchar_\sub@chi\latexchi}
\newcommand{\sub@chi}[2]{%
  \@ifnextchar^{\subsup@chi{#2}}{\latexchi^{}_{#2}}%
}
\newcommand{\subsup@chi}[3]{%
  \latexchi_{#1}^{#3}%
}
\makeatother

\newcommand{\p}{\partial}

\newcommand{\Z}{\mathbb{Z}}

\newcommand{\Lagr}{\mathcal{L}}

\newcommand{\dd}{\mathrm{d}}
\newcommand{\ton}{t_{\mathrm{on}}}

\definecolor{yscolor}{rgb}{0.0, 0.6, 0.0}
\definecolor{yscolor2}{rgb}{0.0, 0.0, 1.0}

\definecolor{cccolor}{rgb}{0.851, 0.467, 0.341}

\definecolor{bgcolor}{rgb}{0.8, 0.0, 0.0}

\begin{document}

\title{Prospects of Detecting Cosmological Domain Walls\\ in the Cosmic Microwave Background}%

\author{Berkin G\"urb\"uz}
\thanks{berkin.gurbuz@mail.mcgill.ca}
\affiliation{Department of Physics, McGill University, Montr\'eal, QC, H3A 2T8, Canada}
\affiliation{Trottier Space Institute, McGill University, Montr\'eal, QC, H3A 2T8, Canada}

\author{Robert Brandenberger}
\affiliation{Department of Physics, McGill University, Montr\'eal, QC, H3A 2T8, Canada}
\affiliation{Trottier Space Institute, McGill University, Montr\'eal, QC, H3A 2T8, Canada}

\author{Yitian Sun}
\affiliation{Department of Physics, McGill University, Montr\'eal, QC, H3A 2T8, Canada}
\affiliation{Trottier Space Institute, McGill University, Montr\'eal, QC, H3A 2T8, Canada}

\begin{abstract}
Topological defects are a generic prediction of field theories with spontaneously broken symmetries, and domain walls, the two-dimensional defects left by a broken discrete symmetry, are among the simplest. We study the imprint of dark, purely gravitationally interacting domain walls on the Cosmic Microwave Background (CMB), working in the regime where a low symmetry-breaking scale $\eta$ allows the walls to evade the Zel'dovich overclosure bound while remaining cosmologically relevant. The relativistic repulsion of matter by a planar wall carves out an evacuated void bounded by an overdense shell, and we compute the resulting photon deflection field analytically for walls both aligned with and arbitrarily inclined to the line of sight. The deflection stamps coherent, strongly non-Gaussian distortions onto the temperature sky, with a characteristic two-peak structure set by the relativistic displacement scale and an inclination dependence that vanishes entirely for the face-on configuration. Because these signatures are localised and anisotropic, they are poorly captured by two-point statistics; we therefore analyse them with a curvelet/wavelet decomposition matched to the wall's geometry, and with simulation-based inference, training a Neural Posterior Estimator (NPE) to recover the scattering redshift $z$, energy scale $\eta$, and inclination $\theta$, and a Neural Ratio Estimator (NRE) to evaluate the likelihood-ratio test statistic for limit-setting. Casting these methods as sensitivity forecasts for a South Pole Telescope-like (SPT) survey, we find that the curvelet and wavelet statistical thresholds both populate a common $\eta \sim 10$--$100\,$MeV band over the accessible redshift range. The neural-ratio sensitivity curve, which exploits the full map morphology, reaches a factor $1.3$--$1.9$ lower in $\eta$ at $z \gtrsim 20$, or $\sim 2$--$7$ in the wall tension, landing essentially on the per-resolution-element noise floor.
\end{abstract}

\maketitle

\section{\label{sec:Intro}Introduction}

Topological defects are a generic prediction of quantum field theories with spontaneously broken symmetries, arising during phase transitions in the early universe \cite{Kibble:1976sj,Vilenkin:2000jqa}. Depending on the topology of the vacuum manifold, these defects may take the form of monopoles, cosmic strings, textures, or domain walls \cite{Vilenkin:1985,Hindmarsh:1994re}. Their formation is governed by the Kibble mechanism \cite{Kibble:1976sj}, and their subsequent evolution provides a potential observational window into high-energy physics inaccessible to terrestrial experiments \cite{Brandenberger:1994db,Brandenberger:2013xyk}. In particular, the constraint of such relics offers a unique probe of symmetry breaking patterns in the early universe, complementary to collider and low-energy searches.

Among these possibilities, domain walls, two dimensional topological defects arising when a discrete symmetry is broken, occupy a special position. Unlike cosmic strings, whose cosmological effects can be subdominant \cite{Hindmarsh:1994re,Brandenberger:1994db}, stable domain wall networks are subject to the overclosure problem: their energy density scales more slowly than radiation or matter and would quickly come to dominate the expansion of the universe if formed at sufficiently high energy scales. This leads to the classic Zel’dovich bound \cite{Zeldovich:1974uw}, which constrains the symmetry-breaking scale associated with domain walls present today to be below the MeV scale in standard cosmological scenarios. However, this apparent problem simultaneously opens an interesting phenomenological window. If domain walls are generated at sufficiently low energy scales, or if their abundance is otherwise suppressed (e.g. by biasing or late-time formation) \cite{Gelmini:1988,Larsson:1996,KibbleLazaridesShafi:1982}, they can evade cosmological bounds while still producing observable imprints. In this regime, domain walls become viable candidates for generating distinct signatures in precision cosmological data.

The cosmic microwave background (CMB) provides a particularly powerful laboratory for probing such relics. As emphasized in studies of cosmic strings \cite{Brandenberger:1994db,Brandenberger:2013,Brandenberger:2013xyk}, topological defects act as active sources of perturbations \cite{Pen:1997ae,Albrecht:1997nt}, continuously sourcing metric fluctuations and leaving non-Gaussian, scale-dependent imprints distinct from those generated by inflationary initial conditions. In the case of domain walls, their planar geometry and relativistic stress-energy tensor lead to characteristic signatures in temperature anisotropies \cite{Sousa:2015,Lazanu:2015}.

Recent advances in observational cosmology significantly enhance the prospects for detecting or constraining such signals. Measurements from the Planck satellite \cite{Planck:2018overview,Planck:2013strings} and ground-based experiments such as ACT \cite{ACT:DR6} and SPT  \cite{SPT3G:2021} have already placed stringent limits on the contribution of topological defects to the CMB anisotropy spectrum \cite{Charnock:2016,Dvorkin:2011aj,Correia:2026}, with no statistically significant detection to date. However, next generation instruments, including the South Pole Telescope-3G, the Simons Observatory, CMB-S4, and LiteBIRD, promise substantial improvements in sensitivity \cite{SPT3G:2021,SimonsObs:2019,CMB-S4:2016,LiteBIRD:2022}. These advances are expected to tighten constraints on defect tensions by factors of a few and may open the possibility of distinguishing between different classes of topological defects through their distinct mode decomposition and non-Gaussian signatures.

Motivated by these developments, it is timely to revisit the observational consequences of domain walls in the CMB. In contrast to cosmic strings, whose signatures often manifest as line-like discontinuities, localized polarization patterns \cite{KaiserStebbins:1984,Danos:2010pol}, or wedge-shaped imprints on the redshifted 21 cm brightness temperature \cite{Brandenberger:2010,Hernandez:2011,Hernandez:2012}, domain walls are expected to induce coherent, planar distortions in the gravitational potential, leading to anisotropic imprints that may be detectable through dedicated statistical analyses. Furthermore, their inherently non-Gaussian nature provides a clear avenue for disentangling their contribution from the dominant primordial fluctuations.

In this work, we investigate the signatures of dark (only gravitationally interacting) domain walls in the CMB, focusing on deriving the non-Gaussian gravitational effects of the domain wall. We emphasize the parameter space in which domain walls evade the overclosure problem while remaining observationally relevant, and we assess the detectability of their imprints in current and upcoming CMB datasets.

In Section \ref{sec:Review}, we give a brief review of topological defect and domain wall physics, and outline the relevant gravitational effects of domain walls. In Section \ref{sec:Profile}, we give a calculation of the density profile that is formed by the repulsive nature of the domain wall. In Section \ref{sec:Deflection}, we give an outline of the calculation done to determine the deflection angle due to the density profile. In Section \ref{sec:CMBDeflection}, we apply this deflection angle to the CMB and highlight key non-Gaussian observables. In Sections \ref{sec:Curvelet} and \ref{sec:SBI} we discuss methods of detecting domain wall induced non-Gaussianities in the CMB by use of curvelet or wavelet analysis and simulation-based inference, respectively. Lastly, we compute the idealized-sensitivity of such methods to detect domain walls in an SPT-3G-like survey in Section \ref{sec:Constraints}.

\section{\label{sec:Review}Domain Wall Review}
Topological defects are generically formed during phase transitions in the very early universe \cite{Kibble:1976sj,Vilenkin:1985}. Unless inflated away, these topological defects persist long enough to leave observable imprints on our cosmology. Such defects commonly arise from spontaneous symmetry breaking in many models that aim to unify the fundamental forces.

Domain walls are a particular class of topological defects, characterized by their two dimensional structure. Domain walls may take planar or spherical forms, which exhibit different dynamics. In this work, we focus strictly on the planar case, as their anisotropic signatures are prominent in the CMB, while their spherical counterparts are generally short-lived by instability. This instability is what allows wall networks to coarsen, allowing the planar signal to stand out.

As a toy model, consider the real scalar field $\phi$ with lagrangian
\begin{equation}
    \Lagr = \frac{1}{2} \p_\mu \phi \p^\mu \phi - \frac{\lambda}{4} \Big( \phi^2 - \eta^2 \Big)^2
    \label{eq:phiLagr}
\end{equation}
with coupling $\lambda$ and symmetry breaking scale $\eta$. Below we give two arguments for the formation of domain walls from this lagrangian. Consider the extrema of the potential
\begin{equation}
    \frac{\p V(\phi)}{\p \phi} = \lambda \phi(\phi^2 - \eta^2) = 0
\end{equation}
The two stable minima are $\phi = \pm \eta$, meaning the vacuum manifold of the model is $\mathcal{M} = \{-\eta, +\eta\} \cong \Z_2$. This lagrangian is invariant under the $\Z_2$ symmetry $\phi \to -\phi$. However, as the temperature of the early universe cools, this symmetry is spontaneously broken. As domain walls are generic consequences of broken $\Z_2$ symmetries, they are inevitably created during the phase transition.

We illustrate this process with our toy model. Suppose that $\phi$ is in thermal equilibrium. Then, at high temperatures, the $\Z_2$ symmetry is restored since $\phi = 0$ is the minimum of the finite temperature effective potential. To see this, note that thermal effects on the field can be encoded with the temperature correction term \cite{Brandenberger:1984cz}
\begin{equation}
    \delta V \sim \lambda' T^2 \phi^2
\end{equation}
For very large $T$, this term dominates the potential, creating a quadratic effective potential. The minima of this effective potential implies that the expectation value $\braket{\phi(x)} = 0$, restoring the symmetry everywhere. As the universe expands and cools, there is a critical temperature $T_c = \eta\sqrt{\lambda / \lambda'}$ where the mass term vanishes. For $T<T_c$, the potential becomes unstable, and the field smoothly rolls down the potential, evolving to one of $\pm \eta$. The direction at which this occurs depends strictly on local thermal and quantum fluctuations which nudge the field in either direction. As the field undergoes the phase transition, the symmetric configuration $\phi = 0$ becomes unstable and the system evolves via spinodal decomposition. Each region of space relaxes into one of the two degenerate vacua, $+\eta$ or $-\eta$. Space is thereby partitioned into domains whose initial size is set by the spinodal scale, the wavelength of the most rapidly growing unstable mode. The domain walls are the two dimensional surfaces that separate adjacent domains of opposite field value. This network subsequently coarsens, with the characteristic domain size growing until it reaches the Hubble scale, yielding the scaling solution \cite{PressRydenSpergel,GaragounisHindmarsh}. Along the domain wall, $\phi \notin \mathcal{M}$, as the field takes neither $\pm \eta$. As such, $V(\phi) > 0$ along the domain wall, hence the domain wall carries energy per unit area, the wall tension. Thus, domain walls are two dimensional regions of trapped energy, capable of affecting structure formation via standard gravitational effects.

A simplified calculation of the gravitational effects of domain walls goes as follows. In the weak field approximation of general relativity \cite{Weinberg:2008,Dodelson:2003} a static matter distribution with energy-momentum tensor $T_\mu^\nu = \text{diag}(\rho, -P_1,-P_2,-P_3)$ obeys the Poisson equation
\begin{equation}
    \nabla^2 \Phi = 4 \pi G (\rho + P_1 + P_2 + P_3)
\end{equation}
A domain wall in the $yz$-plane has a negative pressure along the two dimensions it spans equal to the energy density $|P_2|=|P_3|=\rho$, while $P_1 = 0$. So the wall's energy-momentum tensor can be written
\begin{equation}
    T= \text{diag}(\sigma, 0, \sigma, \sigma) \delta(x)
\end{equation}
where $\sigma$ is the wall tension. Note that a physical domain wall has a thickness $w \sim \lambda^{-1/2} \eta^{-1}$ \cite{Vilenkin:2000jqa}, however, on cosmological scales this is negligible, and we assume the wall to be infinitesimally thin. The corresponding Poisson equation to this energy-momentum tensor is then
\begin{equation}
\nabla^2 \Phi = - 4 \pi G \sigma \delta(x)
\end{equation}
Integrating this once across the domain wall we obtain the familiar result for the repulsive peculiar acceleration
\begin{equation}
    \alpha = - \nabla \Phi = (2 \pi G \sigma) \text{sign}(x)
\end{equation}
We note here that this is a merely simplified calculation, as domain walls do not have static solutions. However, \cite{Vilenkin:1983zq} has shown that plane-symmetric solution to the Einstein equation leads to a metric that is (1+1) dimensional Rindler-like along the time and transverse directions of the domain wall, with proper acceleration $\alpha = 2\pi G \sigma$ \cite{VilenkinGFVDW,Vilenkin:1983zq,Taub:1951}. In either case, the calculation predicts gravitational effects in the form of repulsive acceleration. This repulsion will thus lead to a unique density profile in physical space as matter is pushed away from the domain wall at cosmological times.

Naturally, the scalar field must smoothly evolve from $-\eta$ to $+\eta$ across the domain wall. Solving the equations of motion for the given potential, one obtains
\begin{equation}
    \phi(z) = \eta \tanh\Big( \sqrt{\frac{\lambda}{2}} \eta z \Big) = \eta \tanh\Big(\frac{z}{w}\Big)
\end{equation}
With this, it is easy to determine the tension of the resulting domain wall
\begin{equation}
    \sigma = \int_{-\infty}^\infty dz \ \phi'^2 \sim \sqrt{\lambda} \eta^3
\end{equation}
Crucially, the wall tension goes like $\sigma \sim \eta^3$. This relation highlights why domain walls of high energy theories like the Grand Unified Theory or the Electroweak Theory lead to the overclosure problem \cite{Zeldovich:1974uw,Vilenkin:1985}, the resulting domain walls is highly energetic. This same relation, however, also allows for domain walls as a result of lower symmetry breaking scales to not overclose the universe, granted the symmetry breaking scale $\eta$, and as a result the tension $\sigma$, is low enough.

While the above encompasses the properties of the domain wall, it says very little about why, if at all, topological defects should form. Should a phase transition that subsequently \emph{could} support topological defects in the very early universe occur, the Kibble mechanism \cite{Kibble:1976sj} argues that it is in fact unavoidable that topological defects form. Here we give a simplified argument of the Kibble mechanism focused on the $\mathbb{Z}_2$ symmetry, that once broken, leads to the formation of domain walls.

Once again, the Lagrangian (\ref{eq:phiLagr}) is invariant under the $\mathbb{Z}_2$ symmetry $\phi \to -\phi$. At the critical temperature $T_c$, the field $\phi$ within a patch of the sky must make a choice to settle between $+\eta$ or $-\eta$, but this choice \emph{must} be local. $\phi$ itself is only correlated out to finite distance, the correlation length $\xi(t)$. Thus, two regions of space separated by more than $\xi$ relax into the broken phase independently and can not coordinate their choice in $\mathcal{M}$. Naturally, the field can not be coherent beyond the causal horizon, and so in the radiation era that a domain wall \emph{could} be born, $\xi(t) \lesssim d_H(t) \sim 2t$, where $d_H(t)$ is the causal horizon. Thus, the horizon at the phase transition sets the maximum size a coherent domain of choice in $\mathcal{M}$ can be formed. As a result, the universe is partitioned into patches of such domains, each bounded in size by the horizon. Ultimately, two such domains will meet, and whether a topological defect will be present between the domains is a matter of the topology and disconnectedness of $\mathcal{M}$. For our $\mathbb{Z}_2$ case, each patch may pick between only $+\eta$ and $-\eta$, and it has equal probability of selecting either, $p_+ = p_- = \frac{1}{2}$. Since the domain wall exclusively separates two domains of opposite sign, so that $\phi$ may interpolate between $\pm \eta$ and therefore leave $\mathcal{M}$, the probability that any surface between two domains hosts a domain wall is $p_{\rm wall} = 2 p_+ p_- = \frac{1}{2}$. Therefore, in  a universe divided into patches $\pm \eta$, the boundary between two independently formed domains will host a domain wall half of the time. Note that this high probability simultaneously indicates that domain walls are not formed as a dilute `gas' of isolated sheets, but instead a single, connected network, with mean separation initial $\xi(t) \lesssim 2t$. Importantly, the Kibble mechanism is an argument for the patchiness of space as necessitated by the causal horizon. The topological defects formed are mere consequence of the continuity of the field $\phi$ between these patches. As a result, should the early universe go through a phase transition that breaks a discrete symmetry, such as $\mathbb{Z}_2$, then the formation of domain walls in our universe is \emph{inevitable}.

\section{\label{sec:Signature}Domain Walls in the CMB}
\subsection{\label{sec:Profile}Early Universe Density Profile}
As expressed in Section \ref{sec:Review}, the domain wall act as a repulsive gravitational force perpendicular to its own plane. The strength of this repulsive force on massive particles is determined by the wall tension $G\sigma$ alone, the resulting proper acceleration is $\alpha = 2\pi G \sigma$ \cite{VilenkinGFVDW}. Assuming that the domain wall only begins reshaping matter at matter-radiation equality, when thermal pressure is relieved, matter within the horizon of the domain wall receives a comoving displacement, in direct analogy with the displacement construction used for cosmic string wakes \cite{SilkVilenkin:1984,Vachaspati:1986,Stebbins:1987}, determined by the relativistic equation
\begin{equation}
      \Delta x(q,t)=\int_{\ton(q)}^{t}\frac{u(t';\ton)}{a(t')}\,\dd t'
      \label{eq:Displacement}
    \end{equation}
where $u(t';\ton)$ is the peculiar velocity of matter that entered the gravitational horizon at $\ton(q)$, at time $t'$, where $q$ is the initial comoving position of the element.

For a domain wall formed at $T_c \sim \eta$ or conformal time $\tau_{\rm form}$, its gravitational field begins traveling outwards at the speed of light, so its causal horizon grows $q_H(\tau) = c[\tau - \tau_{form}]$, for conformal time $\tau$. Radiation pressure suppresses structure formation before matter-radiation equality, and so matter does not respond to the domain wall until $\tau_{\rm eq}$. At $t_{\rm eq}$, the gravitational field has already grown to comoving size $q_{\rm eq} = c[\tau_{\rm eq} - \tau_{\rm form}]$. Thus, at $\tau_{\rm eq}$, all matter with $|q| < q_{\rm eq}$ is within the gravitational field of the domain wall, and simultaneously experiences the repulsive acceleration. For this matter, $\ton(q) = t_{\rm eq}$, for cosmic time $t$.

For matter initially outside the causal horizon at $\tau_{\rm eq}$, so $|q| > q_{\rm eq}$, the activation time is determined by the causal horizon crossing $\tau(\ton) = \tau_{\rm form} + q/c$. It can be shown that this is equal to cosmic time
\begin{equation}
    \ton(q)\simeq t_{\rm eq} \left( \frac{q/c + t_{\rm eq}}{3 t_{\rm eq}} \right)^3
\end{equation}
 a result of the $\sim 3ct$ growth of the horizon. In general, the peculiar velocity can be written in terms of the celerity $w$ in the following way
\begin{equation}
    u = \frac{p/m}{\sqrt{1+(p/mc)^{2}}}
        = \frac{c\,w}{\sqrt{c^{2}+w^{2}}}
\end{equation}
where the celerity $w = p/m$ is equivalently the proper velocity. Once again in matter domination, under a constant proper acceleration $\alpha$ due to the domain wall
\begin{eqnarray}
     w(t';\ton) &=& \frac{\alpha}{a(t')}\int_{\ton}^{t'} a(t'')\,\dd t'' = \frac{\alpha}{a(t')} \int_{\tau_{\rm on}}^\tau a^2(\tau')d\tau' \nonumber\\
     &=& \frac{3\alpha t_{\rm eq}}{5\,a(t')}
        \!\left[\left(\frac{t'}{t_{\rm eq}}\right)^{5/3}-\left(\frac{\ton}{t_{\rm eq}}\right)^{5/3}\right]
\end{eqnarray}
Then, an element whose initial comoving position was $q$ at $\ton(q)$, has comoving and physical position at some $t \geq \ton$ equal to
\begin{equation}
    x(q,t)=q+\Delta x(q,t)
      \qquad
      r(q,t)=a(t)\,x(q,t)
\end{equation}
Here we also define $\psi(t) \equiv a(t) \Delta x(0,t)$, which is the physical distance that an element initially adjacent to the domain wall has been repelled out to at $t$. Since the repulsive effect of the domain wall clears out matter towards its transverse direction, $\psi(t)$ defines the bounds of the void region created by the domain wall, extending out to $\pm \psi(t)$.
    
The resulting matter density profile associated to this peculiar velocity field can be determined by conserving the comoving mass along the transverse direction $\rho_{\rm com}(x)dx = \rho_{\mathrm{com},0}dq$. Since the comoving background density is uniform, we utilize the Jacobian $J$ to determine the physical density
\begin{equation}
    \frac{\rho(x,t)}{\bar{\rho}(t)} = \frac{1}{|J|}, \qquad J = \frac{\p x}{\p q}
\end{equation}
The expansion enters only through the background $\bar{\rho}(t)$, while the shape of the density contrast is set entirely by the Jacobian. We can then simply remap to physical coordinates to determine the physical density for all space within the horizon, for all time: $\rho(r,t) = \bar{\rho}(t) / |J|$ evaluated at $r = ax$.

Since $\alpha$ is uniform and common on all elements within the horizon, $q$ dependence of $\Delta x$ only enters through $\ton (q)$, and $d\ton / dq = a(\ton)$. The Jacobian can be written
\begin{equation}
    J = \frac{\p x}{\p q} = 1 + \frac{\p \Delta x}{\p q} = 1 + \frac{\p \Delta x}{\p \ton}\frac{d \ton}{dq}
\end{equation}
Upon integrating by parts and dropping the boundary term, it can be shown that
\begin{equation}
    \frac{\p \Delta x}{\p q} = a(\ton)\frac{\p \Delta x}{\p \ton} = a(\ton)\int_{t_{\rm on}}^{t}\frac{1}{a(t')}\frac{\partial u}{\partial t_{\rm on}}\,dt'
\end{equation}
Then, writing $\frac{\p u}{\p \ton} = \frac{d u}{d w} \frac{\p w}{\p \ton}$ where
\begin{equation}
      \frac{\partial w}{\partial\ton}=-\frac{\alpha\,a(\ton)}{a(t')},
      \qquad
      \frac{\dd u}{\dd w}=\frac{1}{\big[1+(w/c)^{2}\big]^{3/2}},
    \end{equation}
We can finally assemble the Jacobian to be
\begin{equation}
      J(q,t)=1-\frac{\alpha\,a(\ton)^{2}}{c}
      \int_{\ton}^{t}\frac{\dd t'}{a(t')^{2}\,\big[1+(w(t';\ton)/c)^{2}\big]^{3/2}}
      \label{eq:jacobian}
    \end{equation}
The fraction $\rho / \bar \rho = 1 / |J|$ is single-valued as long as $J>0$. Shell crossing would occur when $J=0$, causing neighboring fluid elements to overlap and producing a caustic. In this fully relativistic setup, the speed limit on the local peculiar velocity ($u<c$) prevents shell crossing from ever occurring, which we also check numerically. As a result, caustics never form, and the resulting density profile can be used at all redshifts that we probe. We further assume that the domain wall persists until the redshift at which it is observed, continuously sourcing its repulsive gravitational field.

\begin{figure}
    \centering
    \includegraphics[width=1\linewidth]{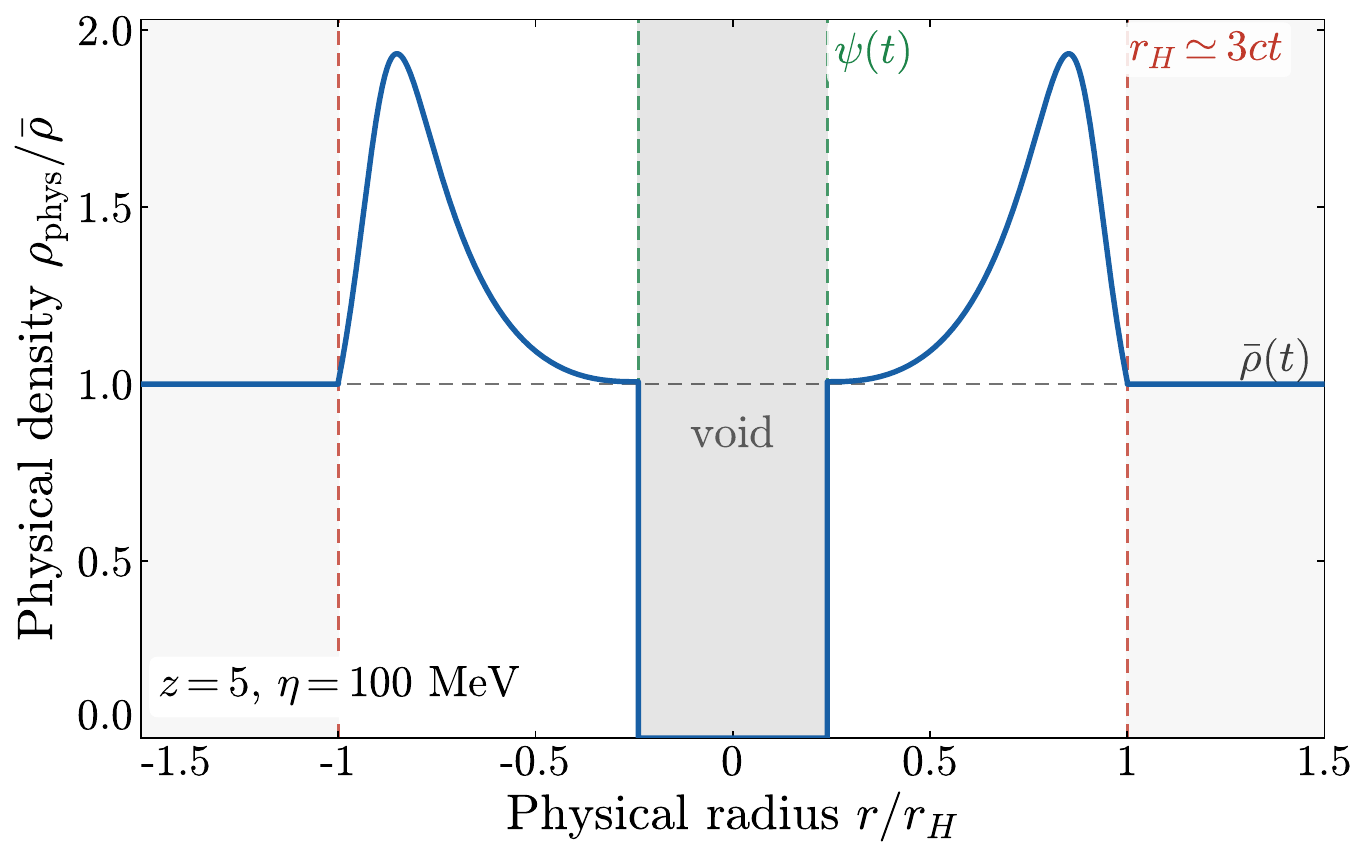}
    \caption{Density profile created by the repulsion of the domain wall. Gray shaded region indicates the evacuated void region. Black, red, and green dashed lines represent the background physical density, matter-era horizon, and void boundary $\psi(t)$, respectively.
    }
    \label{fig:Profile}
\end{figure}

\begin{figure}
    \centering
    \includegraphics[width=\linewidth]{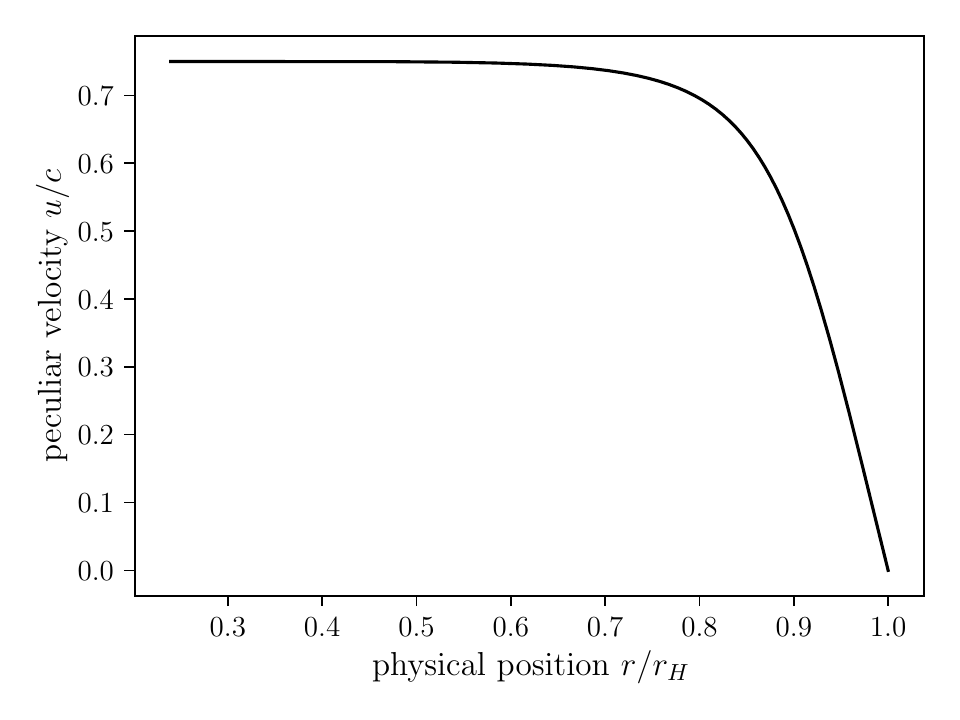}
    \caption{Peculiar velocity $u$ field of matter between the void edge $+\psi(t)$ and the horizon $+r_H(t)$. The density profile Figure \ref{fig:Profile} is a direct consequence of the derivative $du/dr$ near $\psi$ and $r_H$.}
    \label{fig:Peculiaru}
\end{figure}

Given in Figure \ref{fig:Profile}, as expected our profile has an underdense void region bounded by $\psi(t)$, followed by an overdense shell that begins at the background, peaks shortly before the horizon, and converges back onto the background at the horizon. We note here that the formation of the overdense region is not due to self-gravity of the matter, but the fact that matter only experiences the repulsive field when within the horizon of the domain wall. Two comoving-adjacent elements $q$ and $dq$ are reached by the walls light cone at cosmic time separation $d\ton = a(\ton)dq$. This separation is the headstart that the inner element gets to pull closer to the outer one. Near the wall, however, $a(\ton) \simeq 1$ and so neighboring elements switch on almost simultaneously and stream together with roughly equal velocity. These elements, despite having the greatest peculiar velocity, have relatively low shear and compression. Near the horizon, however, $a(\ton)$ is quite large, and successive neighbors are \emph{activated} at greater and greater time separations. This accumulated headstart produces a strong compression, as inner elements catch up to outer elements. In fact, the compression grows outwards, and weighted by the $a(\ton)^2$ in the Jacobian, and suppressed only at the horizon, where freshly \emph{activated} elements have not had enough time to move much yet. As a consequence, the peak of the overdensity sits just below the horizon on either side of the domain wall. Interestingly, the inner-most matter at the void edge $\psi(t)$ is also the fastest matter. Despite this, since the peculiar velocity field is almost constant in this region, the density too remains uniform. 
This is verified in Figure \ref{fig:Peculiaru}.

The transverse extent of the wall that enters the deflection integrals is fixed by the cosmological evolution of the wall network. At formation, the spinodal decomposition described above partitions space into domains whose initial size is set by the wavelength of the fastest-growing unstable mode, comparable to the wall thickness $w \sim (\sqrt{\lambda}\,\eta)^{-1}$, and therefore microscopic on cosmological scales. This network is far from equilibrium. The wall tension acts to minimize area, so the walls straighten, collide and reconnect, and pinch off closed surfaces that subsequently collapse, with the lost energy predominantly carried away by scalar radiation \cite{GaragounisHindmarsh}. The characteristic scale of the network grows under this coarsening until it reaches the only macroscopic length available to it, the Hubble radius.

Beyond this point, the network relaxes onto the well-known \emph{scaling solution}, a dynamical attractor in which the wall curvature radius and the inter-wall separation both track the horizon
\begin{equation}
    R \sim L \sim H^{-1} \sim t,
    \label{eq:wallscaling}
\end{equation}
so that the comoving wall area density falls like $A/V \propto t^{-1}$ and the network on average maintains one wall per Hubble volume \cite{PressRydenSpergel,GaragounisHindmarsh,Hiramatsu}. The corresponding energy density, $\rho_{\rm wall}\sim \sigma/L \sim \sigma H$, redshifts only as $t^{-1}$, more slowly than matter or radiation, which is precisely the behaviour underlying the Zel'dovich overclosure bound of Section \ref{sec:Review}. The wall scale and the Hubble radius are proportional up to some $\mathcal{O}(1)$ constant, the area parameter $A$, whose value has been measured in field-theoretic simulations of $\mathbb{Z}_2$ wall networks to be $\mathcal{A}\simeq 0.8\pm0.1$ \cite{Hiramatsu,Saikawa}. We adopt this value to fix the coherent extent of a single wall at the redshift of scattering
\begin{equation}
    L(z) = \frac{\mathcal{A}}{H(z)}, \qquad \mathcal{A}\simeq 0.8 .
    \label{eq:Lz}
\end{equation}
In absolute terms this scale is enormous, growing from tens of $\mathrm{Mpc}$ to nearly a $\mathrm{Gpc}$ across the redshift range we are interested in, but this is simply the statement that the wall fills a fixed fraction of the horizon, as obtained in the scaling solution.

The scaling regime is reached within a few Hubble times of formation, and for the low symmetry-breaking scales of interest $\eta \sim \mathcal{O}(1-100)\,\mathrm{MeV}$, the wall-forming transition occurs at $z \sim 10^{11}$, many expansion times before the epochs $z \lesssim 100$ that we probe. Therefore, any memory of the microscopic spinodal scale is long erased, and over our entire range the network firmly sits at the scaling regime where (\ref{eq:Lz}) applies. Since $L$ is set by the horizon, we treat each domain wall as a planar sheet of finite extent $L \times L$ given by (\ref{eq:Lz}). This finite extent is what renders the line-of-sight and transverse integrals of the deflection in Section \ref{sec:Deflection} to be convergent. At the lowest redshift in our range, $L$ exceeds the $10^\circ$ field of view, and the wall behaves effectively infinite across our observational window, while at higher redshift the finite size remains resolvable.

\subsection{\label{sec:Deflection}Photon Deflection}
The density profile given at the end of Section \ref{sec:Profile} will naturally cause incident photons to be deflected. Given the planar symmetry of our system, the resulting deflection angle should be antisymmetric as we observe opposite sides of the domain wall.

A photon whose light ray is parameterized by $(\vec{\zeta}(\lambda),z(\lambda))$, where we have chosen the line of sight and propagation direction to be $z$, is deflected by the under and over dense regions of the density profile. Utilizing the Born scattering approximation \cite{Bartelmann:1999}, we can simplify $\vec{\zeta}(\lambda) = \vec{\zeta}$ to be a constant vector. Then, the resulting deflection to this photon is given by
\begin{equation}
    \vec{\alpha} = \frac{4G}{c^2} \int d^2 \vec{\zeta'} \underbrace{\int dz' \left[\rho(\vec{\zeta'},z',t) - \bar{\rho}(t)\right]}_{\equiv \Sigma(\vec{\zeta'},t)} \frac{\vec{\zeta} - \vec{\zeta'}}{|\vec{\zeta} - \vec{\zeta'}|^2}
\end{equation}
\begin{figure}[t]
    \centering
    \includegraphics[width=1\linewidth]{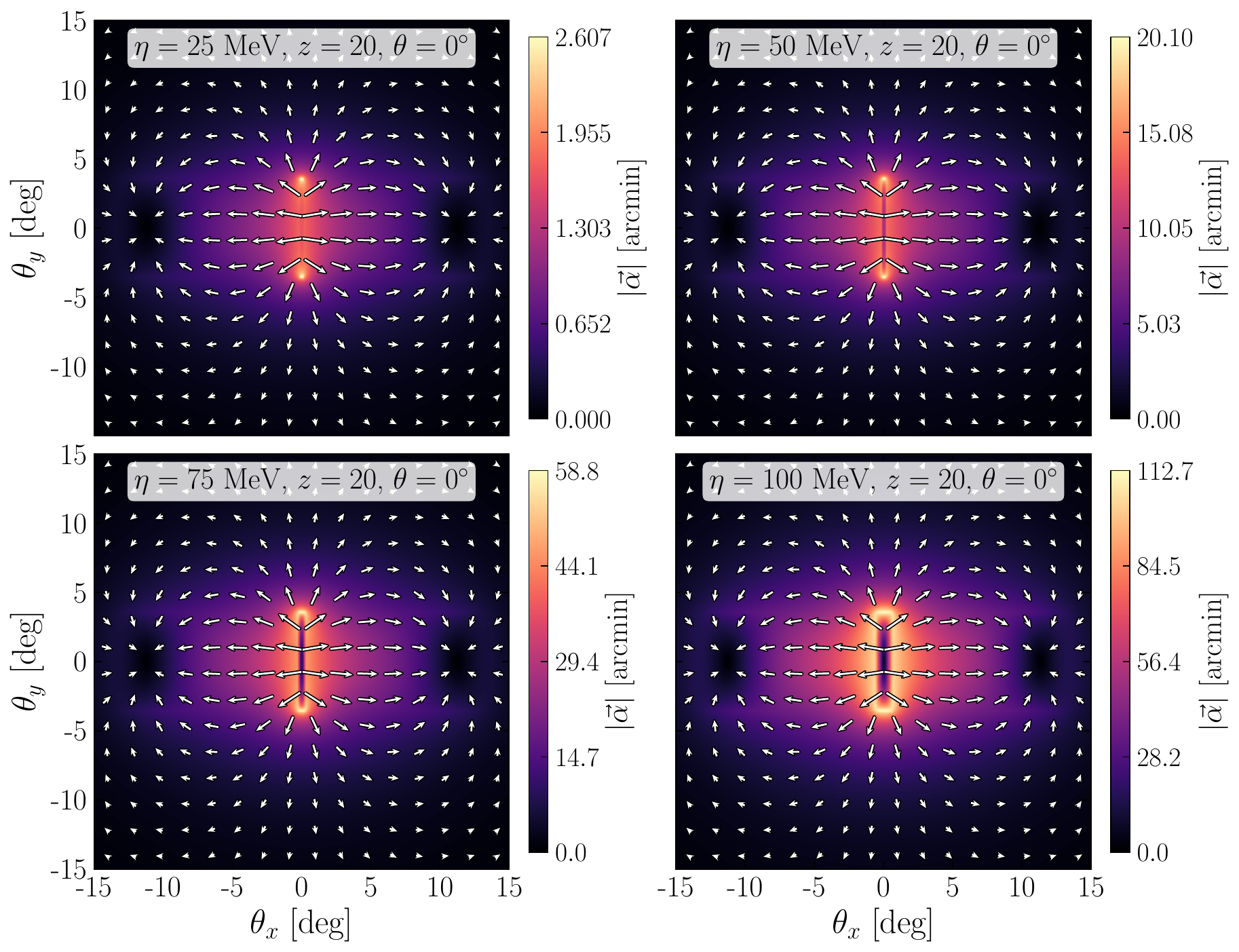}
    \caption{Deflection field $\vec \alpha(\vec \theta)$ of photons due to the matter profile around the domain wall. Each window has subsequently increasing value for $\eta$, at fixed scattering redshift $z = 20$ and inclination $\theta = 0^\circ$. Arrow orientation indicates the deflection direction, while the arrow length scales with local deflection strength, also indicated by the colormap. While the deflection field is generated by the matter profile, we note that in each simulation the domain wall is centered at $\theta_x=0$, with extent $\theta_y \in [-L(z)/2,L(z)/2]$ and $L(z)$ along the line of sight.}
    \label{fig:Deflection}
\end{figure}
 where $\Sigma$ is line-of-sight projected surface density contrast. Note that only a statistically insignificant number of domain walls would be perfectly lined up with our line of sight. The correct approach is then to consider domain walls with an arbitrary orientation in space. Specifically, we consider a domain wall rotated by an inclination angle $\theta$ around the $y$ axis, perpendicular to the line of sight $z$. Any rotation around the $z$ axis leads to degenerate results as the observational window can simply be rotated, not leading to any new physics. For a domain wall, and as a result the matter profile, rotating by an inclination $\theta$, we utilize the rotated coordinates
 \begin{equation}
  u = x'\cos\theta - z'\sin\theta,\qquad w = x'\sin\theta + z'\cos\theta,
\end{equation}

\begin{figure}[t]
    \centering
    \includegraphics[width=1\linewidth]{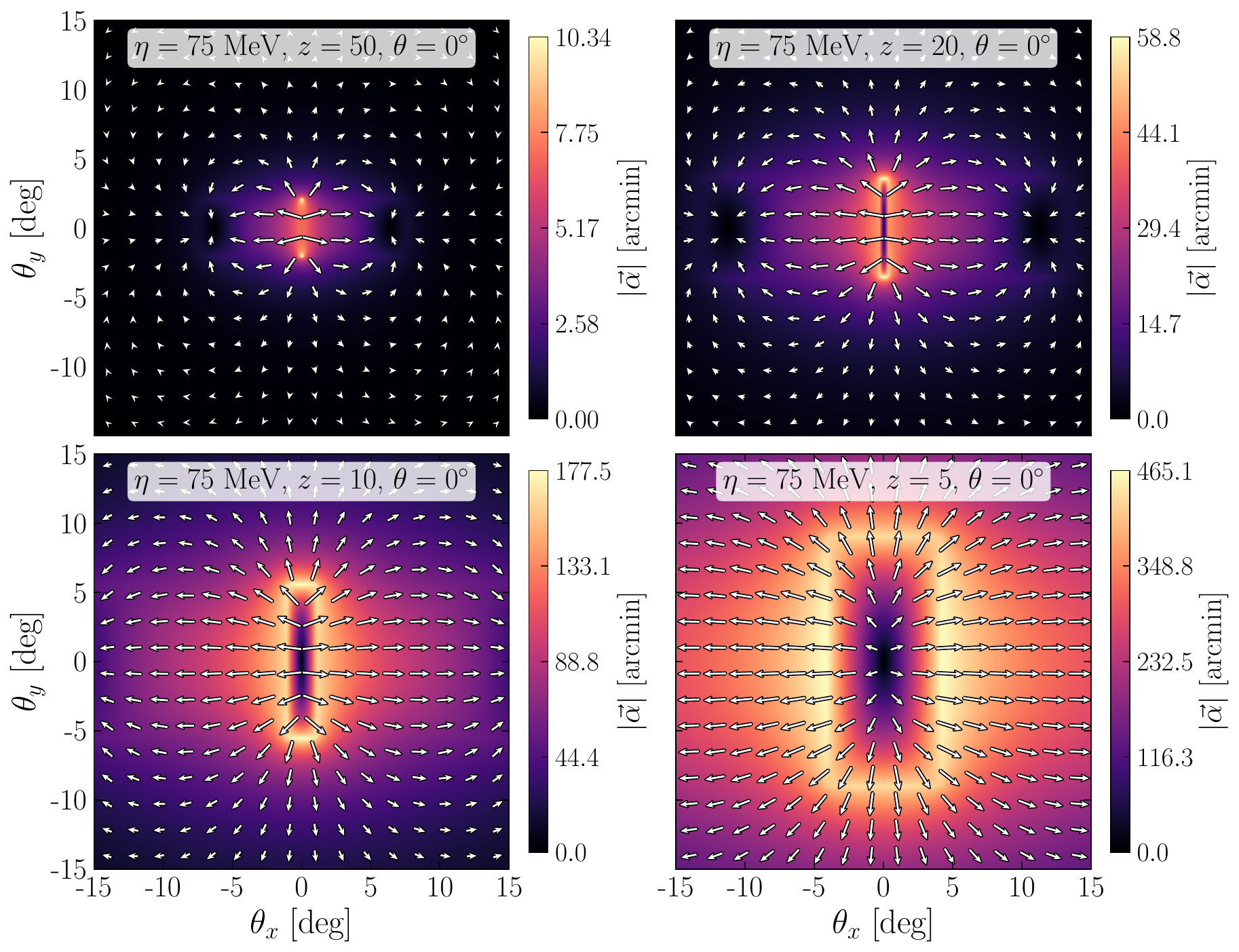}
    \caption{
    Deflection field $\vec \alpha(\vec \theta)$ of photons due to the matter profile around the domain wall. Each window has subsequently increasing value for scattering redshift $z$, at fixed scale $\eta = 75 \rm MeV$ and inclination $\theta = 0^\circ$. Arrow orientation indicates the deflection direction, while the arrow length scales with local deflection strength, also indicated by the colormap.}
    \label{fig:DeflectionZ}
\end{figure}

\begin{figure}
    \centering
    \includegraphics[width= \linewidth]{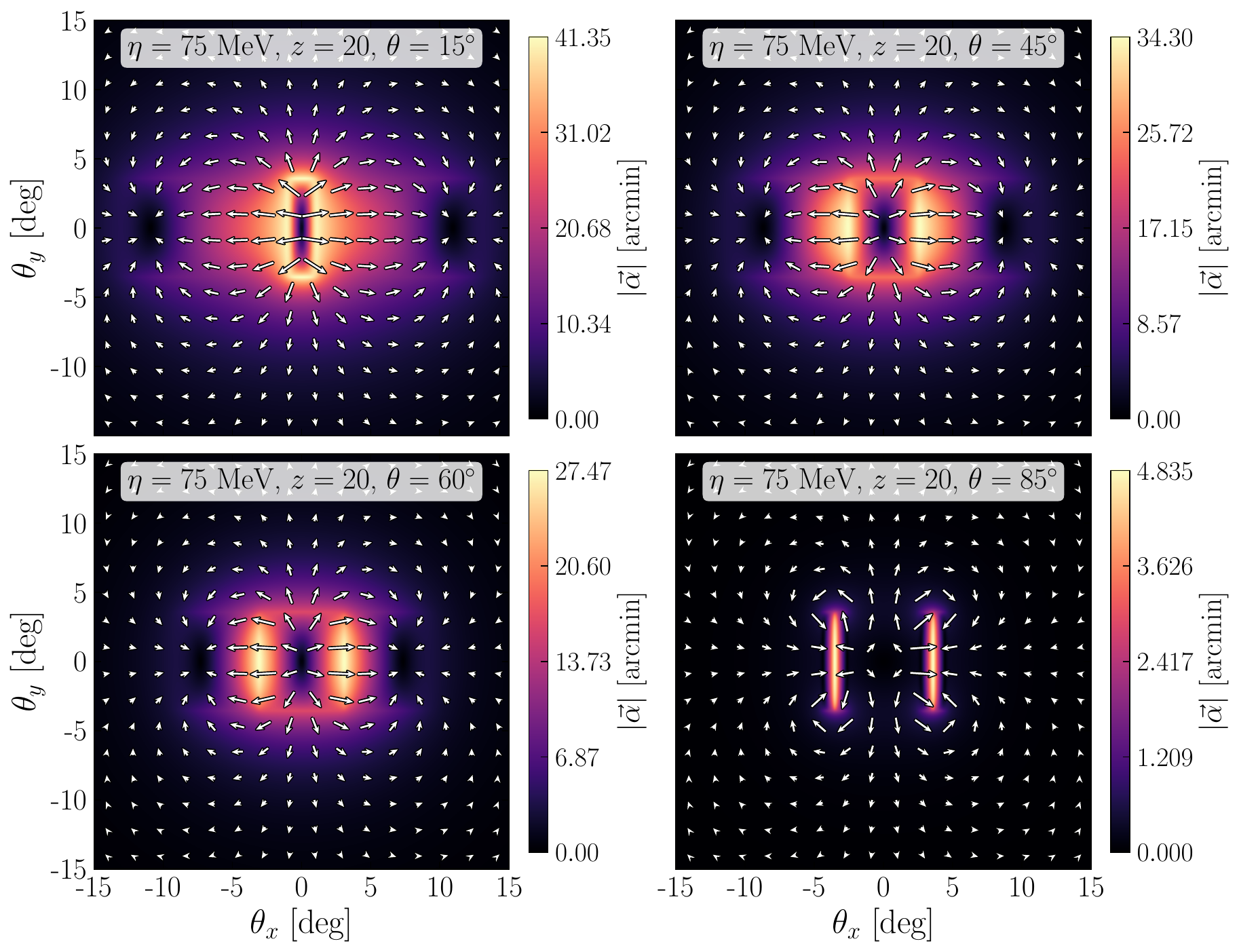}
    \caption{
    Deflection field $\vec \alpha(\vec \theta)$ of photons due to the matter profile around rotated domain walls. $\theta$ indicates rotation of the domain wall plane around the $y_0$ axis of the plane. $\theta$ is an inclination angle such that $\theta=0$ corresponds to the perfectly aligned wall as analyzed in Figure \ref{fig:Deflection}, while $\theta = 90^\circ$ is the ``face-on" wall, where the wall normal is aligned with the line-of-sight, and the deflection vanishes. Each window is evaluated at the fixed scale $\eta = 75 \rm MeV$ and scattering redshift $z = 20$.}
    \label{fig:DeflectionT}
\end{figure}

Then, since the density contrast depends only on the distance perpendicular to the domain wall, $\Sigma = \Sigma(u)$, while the finite extent of the domain wall will confine $w,y'\in[-L/2,L/2]$. Thus, the $w$ and $y'$ integrals in the Born kernel act only on the geometry and not on the density through $\Sigma$. As a result, these two can be carried out analytically, yielding
\begin{equation}
  \alpha_{x,y}(x_0, y_0)=\frac{4G}{c^{2}}\int_{0}^{r_H}\!du\;\Sigma(u)\,
  \bigl[K_{x,y}(u)+K_{x,y}(-u)\bigr]
  \label{eq:alpha_quadrature}
\end{equation}
where $K_{x,y}(u)$ is a geometric kernel, whose full definition and derivation can be found in Appendix \ref{sec:AppDeflection}, fixed solely by the inclination and wall extent. The folding $K(u) + K(-u)$ exploits the fact that $\Sigma(u)$ is even in $u$. This integral of the fully relativistic FLRW matter profile must be computed numerically, and we do so by use of Gauss-Legendre quadrature. The results for a range of scale $\eta$, scattering redshift $z$, and inclination $\theta$ can be found in Figures \ref{fig:Deflection}, \ref{fig:DeflectionZ}, \ref{fig:DeflectionT}. We take the convention that $\alpha_x < 0$ is a right-ward deflection of photons, while $\alpha_x > 0$ is a left-ward deflection. This suggests that the matter profile drags photon paths away from the domain wall in the transverse direction. Therefore, the domain wall behaves like a concave, or diverging, optical lens.

We highlight a few characteristics of the deflection field: just as the domain wall's own gravitational field behaves repulsively, we observe that the resulting matter profile also deflects photons away from the domain wall central to the void region. The arrows in Figures \ref{fig:Deflection}, \ref{fig:DeflectionZ}, \ref{fig:DeflectionT}, which demonstrate the deflection direction, shows that photons are pushed away from the void region, and pulled in towards the overdensity. Evidently, the boundary of the void region is where the deflection field is at its strongest.

Raising the symmetry breaking scale $\eta$ and lowering the scattering redshift $z$ both strengthen the deflection field. The former is understood by the repulsive acceleration strength $\alpha = 2\pi G \sigma$, where the tension $\sigma \sim \eta^3$ grows very fast with the scale. A stronger repulsive field leads to a larger evacuated void region and overdensity peak, both of which contribute to the strength of the deflection field. Similarly, the latter is understood by recalling that the domain wall and matter profile are both continuously evolving in time. The domain wall extent continues to evolve on the scaling regime $L\sim t$, while the matter profile has more time to evolve under the repulsive field, culminating in a larger void region and greater overdensity.

The growing deflection strength suggests some degeneracy between the effects of raising the scale $\eta$ and lowering the scattering redshift $z$. A simple way to distinguish the two, however, is to look at two attractive regions on either side of the void where the overdense region resides. As seen best in Figure \ref{fig:Deflection}, raising the scale $\eta$ does not push the two attractors outwards, although it does cause the void to grow. This is understood by recalling that the location of the overdesity in the matter profile given in Figure \ref{fig:Profile} is primarily determined by the physical causal horizon $r_H$, rather than the repulsion. This factor is completely independent of the repulsive field, and as a result of $\eta$. In contrast, a lower scattering redshift $z$ is directly sampling a time when the causal horizon $r_H$ has grown, dragging the overdensity along with it. Lastly, we note that the wall extent $L$ depends only on the time that the domain wall has had to evolve, and not on $\eta$. In Figure \ref{fig:DeflectionZ} it can be seen that the extent of the deflection signal along $\theta_y$ grows with lowered redshift, while in Figure \ref{fig:Deflection}, \ref{fig:DeflectionT}, a constant wall extent implies a constant signal extent along $\theta_y$.

\subsection{\label{sec:CMBDeflection}Signatures in the CMB}

\begin{figure}
\centering
\includegraphics[width =1\linewidth]{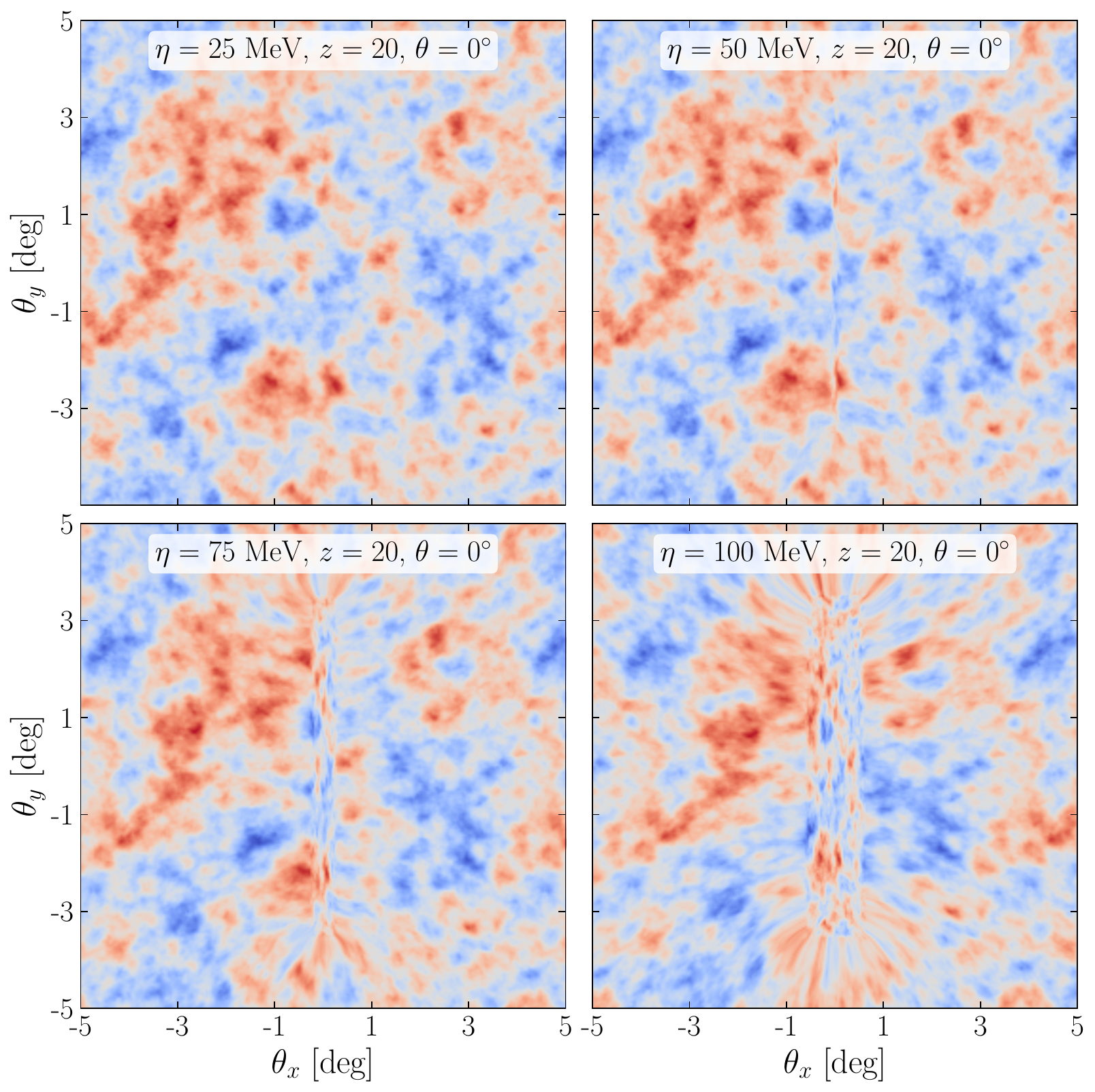}
\caption{CMB maps with Planck-normalized \cite{Planck:2018} $\Lambda$CDM fluctuations deflected by domain walls of growing wall tension $\sigma$. The walls are taken to be centered at $(\theta_x,\theta_y) = (0,0)$ on a $10^\circ \times 10^\circ$ observational window. Stronger domain walls create strong shears in the temperature field easily distinguishable by eye. All of these domain walls began clustering at $t_\mathrm{eq}$ and the observed photons were scattered at $z = 20$.}
\label{fig:CMBDef}
\end{figure}

This analysis assumes that domain walls begin to gravitationally repel surrounding matter at the time of matter-radiation equality. The scaling solution of domain walls implies that there should be at least one domain wall per Hubble volume. This gives us a distribution of domain walls throughout the universe. Photons of the CMB can then scatter off these domain walls at various redshifts, leading to a signature. Thus, not only does the domain wall tension and orientation in space affect the signature, but also the time elapsed after matter-radiation equality, here given as the redshift, leads to vastly varying signatures.

When utilizing the deflection angle and making observations today, it is crucial to make use of the lensing efficiency factor \cite{Bartelmann:1999,LewisChallinor:2006}
\begin{equation}
    \frac{D_{\ell s}}{D_{s}} = 1 - \frac{\chi_{\scriptstyle{\ell}}(z)}{\chi_s} = 1 - \frac{\chi_\ell(z)}{\chi_{\rm CMB}}
\end{equation}
where $\chi$ represents the comoving distance and subscripts $\ell$ and $s$ correspond to lens and source, respectively. Naturally, the source is taken to be the CMB, while the $\chi_\ell$ is the comoving distance to the scattering redshift $z$. Then, observed deflection field $\alpha_{\rm obs}$ is determined by wall-frame deflection $\alpha$ by use of the efficiency factor \cite{Schneider1992}
\begin{equation}
    \alpha_\mathrm{obs}(\theta) = \frac{D_{\ell s}}{D_{s}} \alpha (\theta)
\end{equation}
The deflection angles given in Section \ref{sec:Deflection} are applied to a CMB temperature map as a coordinate transformation \cite{LewisChallinor:2006}, with this efficiency factor taken into account
\begin{equation}
    T_{\text{obs}}(\theta) = T_\text{prim}(\theta-\alpha_\mathrm{obs}(\theta))
\end{equation}
where $T_\text{prim}$ is the primary, unlensed CMB. Note that the deflection angle $\alpha(x_0,y_0)$ in Section \ref{sec:Deflection} must be mapped to the angular space of the observational window, $\alpha(\theta_x,\theta_y)$, by use of the angular diameter distance $D_A = a(z) \chi(z)$ via
$$\theta_x = \frac{x_0}{D_A(z)} = \frac{x_0}{a(z) \chi(z)}$$
where $a(z)$ is the scale factor and $\chi(z)$ is the comoving distance to redshift $z$. The lensing effects of various domain walls on the CMB can be found in Figure \ref{fig:CMBDef}. The key feature is that the deflection creates very non-Gaussian line-like imprints on the CMB. These non-Gaussianities make it so that statistical methods are especially useful at extracting the domain wall signal from thermal noise. Specifically, the effect of the domain wall appears to be a linear \emph{pinch} transverse to the wall. Since the computed deflection describes the domain wall to behave like a diverging lens, it is expected that it would have a similar demagnifying property, as observed.

\begin{figure*}
\centering
\includegraphics[width =\linewidth]{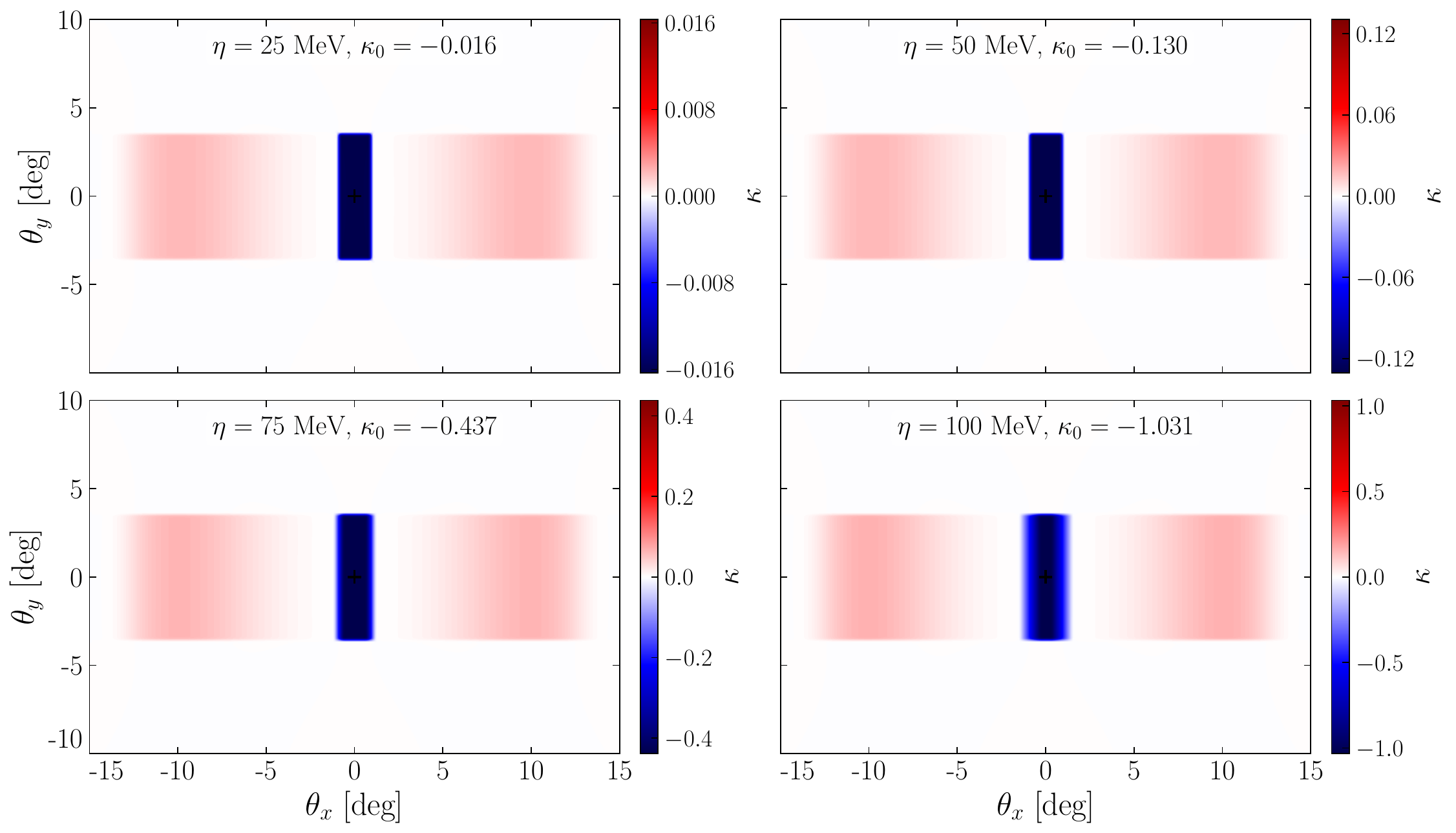}
    \caption{Convergence $\kappa$ of the domain wall deflection field for growing wall tension $\sigma$. The walls are taken to be centered at $(\theta_x,\theta_y) = (0,0)$ on a $10^\circ \times 10^\circ$ observational window. Negative convergence indicates the evacuation, or divergence, of photons, while positive convergence indicates the overlap, or convergence, of photon rays. All of these domain walls began clustering at $t_\mathrm{eq}$ and the observed photons were scattered at $z = 20$ with $\theta=15^\circ$ wall inclination.}
    \label{fig:Convergence}
\end{figure*}

To quantify this demagnification, we make use of the convergence. We begin by considering an initially uniform background of photons, and deflect it. Naturally, the deflection shifts photons around, causing rays to converge or diverge. The convergence is then defined to be
\begin{equation}
    \kappa(\theta) = \frac{1}{2}\left(\frac{\partial \alpha_x}{\partial \theta_x}+\frac{\partial \alpha_y}{\partial \theta_y}\right)
\end{equation}
The convergence comes from the lensing Jacobian of the deflection \cite{Bartelmann:1999}
\begin{equation}
    A_{ij} = \delta_{ij} - \frac{\partial \alpha_i}{\partial \theta_j}
\end{equation}
where the trace-part corresponds to the convergence, while the trace-less part is the shear. We compute $\kappa$ to obtain Figure \ref{fig:Convergence}. The convergence of the domain wall is exceedingly strong. The negative region, due to the evacuation from the domain walls repulsion, is comparable to supervoids \cite{Melchior:2014}. The positive region, likewise, demonstrates strong lensing comparable to galaxy clusters and galactic centers. As expected by the shape of the matter profile, the peak density, and as a result the peak positive convergence lie far away from the domain wall, close to the horizon. The observed negative convergence is a key tell of the demagnification observed in the lensed CMB.

Lastly, we note that the convergence appears as slivers the observational window with sharp cutoffs. This can be easily understood by seeing that the convergence is simultaneously equivalent to
\begin{equation}
    \kappa = \frac{\Sigma(\theta)}{\Sigma_{\text{crit}}}
\end{equation}
where $\Sigma$ is the line-of-sight projected density, and $\Sigma_\mathrm{crit}$ is the line-of-sight project density as a result of the critical density $\rho_c$, which is simply the background $\rho_0$. For regions above the domain wall plane, the density profile converges onto to the background rapidly. As a result, the convergence vanishes. Thus, we also simultaneously find that $\partial_{\theta_x} \alpha_x = - \partial_{\theta_y}\alpha_y$ for $|y_0|>L(z)/2$.

\begin{figure}[t]
    \centering
\includegraphics[width=1\linewidth]{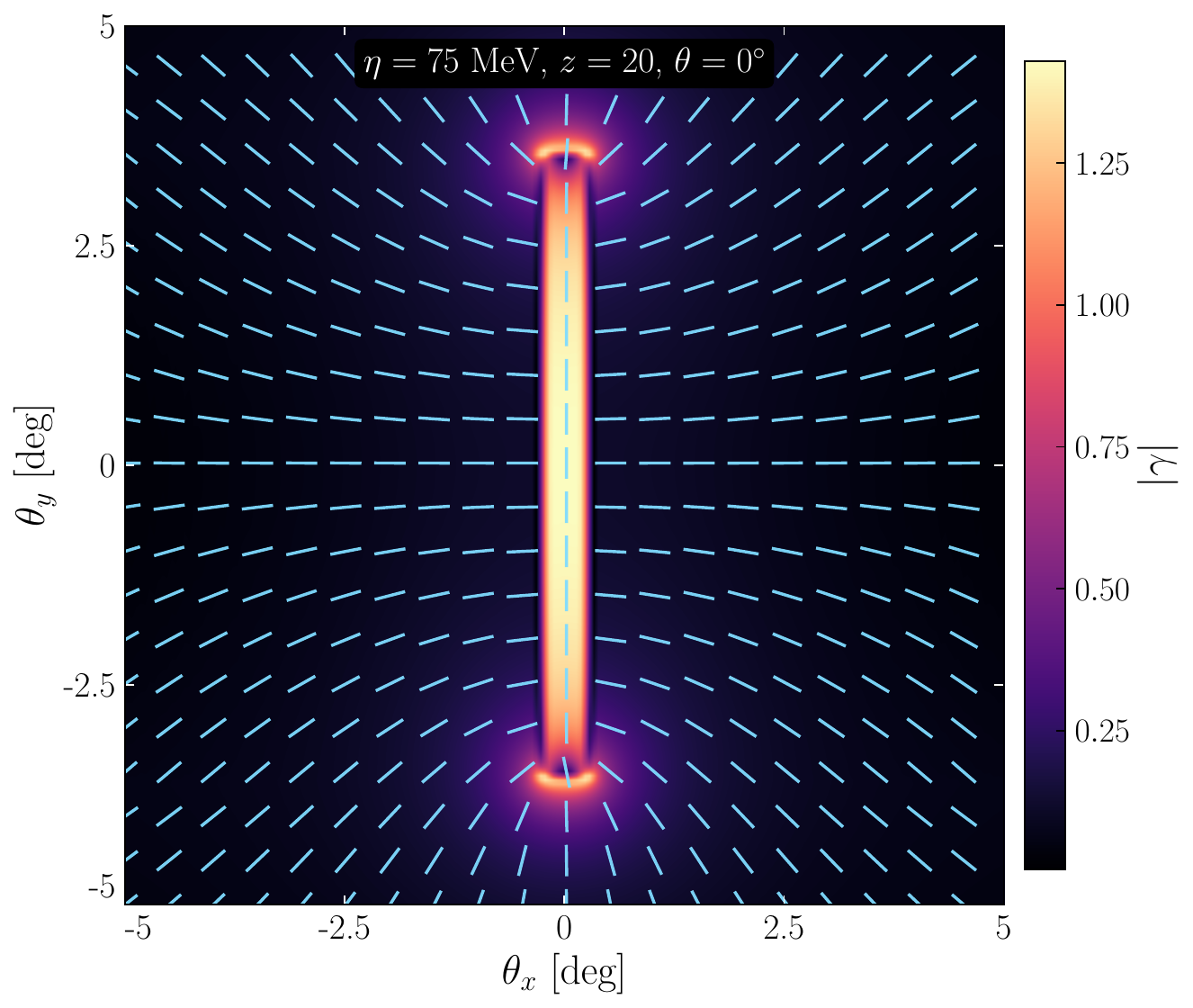}
    \caption{Shear of the lensing due to a domain wall with $\eta = 75\,$MeV,
$\theta = 0^\circ$, $z = 20$. Whiskers are of fixed length and indicate only
the orientation $\varphi = \tfrac{1}{2}\,\mathrm{atan2}(\gamma_2,\gamma_1)$,
while the colour gives the local magnitude $|\gamma|$. Along the wall the
whiskers lie parallel to the wall plane, the signature of the transverse
pinch imprinted on the temperature field.}
    \label{fig:Shear}
\end{figure}

The trace-less part of the lensing Jacobian is the shear $\gamma = \gamma_1 + i \gamma_2$ where $\gamma_1 = \frac{1}{2}(\p_x \alpha_x - \p_y \alpha_y)$ and $\gamma_2 = \frac{1}{2}(\p_y \alpha_x + \p_x \alpha_y)$. Whereas the convergence rescales (magnifies or demagnifies) an image, the shear distorts its shape by mapping a circular source into an ellipse whose ellipticity is $|\gamma|$ and major axis along the direction $\phi = \frac{1}{2} \mathrm{atan2}(\gamma_2, \gamma_1)$. The length and direction of the whiskers in Figure \ref{fig:Shear} represent the local shear magnitude $|\gamma|$ and the orientation $\phi$ of distortions. Along the wall, the whiskers lay parallel to the wall. This describes the observed \emph{pinching} in the lensed CMB.

These results confirm what was observed in the lensed CMB. The domain wall acts exactly like a diverging lens, and the CMB image is greatly demagnified and distorted along the domain wall plane. While the demagnification is comparable to supervoids, the planar distortion is a signal unique to domain walls.

\section{\label{sec:Curvelet}Curvelets and Wavelets}
The gravitational lensing signatures produced by cosmological domain walls are strongly non Gaussian and spatially localized. In particular, the planar geometry of the wall creates coherent line like distortions in the observed CMB temperature field. Such features are not efficiently characterized by standard two point statistics such as the angular power spectrum $C(\ell)$ \cite{HuDodelson:2002}, since these observables are primarily sensitive to Gaussian and statistically isotropic fluctuations. As a result, statistical methods capable of identifying localized anisotropic structure are especially useful for detecting domain wall signatures.

    Wavelet transforms provide one such framework \cite{Daubechies:1992}. Unlike Fourier methods, which only encode global frequency information, wavelets simultaneously localize information in both angular position and scale. Given a CMB temperature map $T(\hat{n})$, the wavelet transform projects the map onto a family of basis functions $\psi_{j,\bm{x}}$ characterized by a scale $j$ and spatial position $\bm{x}$. The resulting wavelet coefficients are given by
$$W(j,\bm{x}) = \int d \Omega \, T(\hat{n}) \psi_{j,\bm{x}}(\hat{n})$$
These coefficients measure localized fluctuations of the temperature field at different angular scales. Compact non-Gaussian structures therefore become enhanced relative to the approximately Gaussian primordial CMB background.

In the case of domain walls, the induced deflection field creates coherent temperature gradients concentrated near the boundaries of the evacuated void and overdense shell. These gradients naturally produce excess power in wavelet coefficients at angular scales corresponding to the characteristic size of the wall profile. Since the primordial CMB is approximately Gaussian, the coefficient distribution for standard $\Lambda$CDM maps is also close to Gaussian. The presence of a domain wall therefore manifests as localized outliers and heavy tails in the coefficient distribution.

While ordinary wavelets are effective at detecting compact localized structures, curvelets are especially useful for identifying elongated anisotropic features. Curvelets form a directional multiscale basis optimized for representing curves and edges \cite{Starck:2002}. Their anisotropic scaling relation $\text{width} \sim (\text{length})^2$ makes them highly sensitive to coherent line like structures such as those produced by topological defects. This property has previously proven successful in searches for cosmic string signatures in the CMB \cite{Hergt:2017,Stewart:2009}, where the Kaiser Stebbins effect \cite{KaiserStebbins:1984} similarly generates localized discontinuities.

Applying a curvelet decomposition to simulated lensed CMB maps therefore provides a natural method for isolating domain wall signals. In practice, the observed temperature field is decomposed into curvelet coefficients $C_{b,k}(\mathbf{x})$, where $b$ labels the radial frequency band, $k$ labels the orientation, and $\mathbf{x}$ labels spatial position. Domain wall signatures are expected to preferentially populate a narrow range of scales and orientations corresponding to the projected geometry of the wall. Statistical estimators constructed from the coefficient amplitudes, kurtosis, peak counts, or higher order moments can then be used to distinguish domain wall realizations from purely Gaussian CMB maps \cite{Hergt:2017,Amsel:2008,Danos:2010,McDonough:2013}. An additional advantage of multiscale methods is that they naturally separate coherent physical structure from instrumental noise: the domain wall signal is spatially correlated while detector noise is uncorrelated, so that wavelet and curvelet filtering can substantially improve signal to noise even when the wall occupies only a small fraction of the observational window.

To make the directional sensitivity of the curvelet basis concrete, it is
useful to collapse the full set of coefficients into a single spatial map. At a
fixed band $b$, the decomposition returns one coefficient array $C_{b,k}(\mathbf{x})$
per orientation $k$, each responding most strongly to edges aligned with
that wedge. Since the projected orientation of a given wall is not known a
priori, we marginalize over orientation by summing the coefficient magnitudes
in quadrature, defining the curvelet amplitude map
\begin{equation}
    A_b(\mathbf{x}) = \left[ \sum_{k} \left| C_{b,k}(\mathbf{x}) \right|^2 \right]^{1/2}.
    \label{eq:curvelet-amplitude}
\end{equation}
This quantity is non-negative and rotationally pooled: it measures the total
directional power present at position $\mathbf{x}$ and band $b$, regardless of which way
an edge happens to point. A coherent line-like feature deposits power into the
handful of wedges aligned with its normal, so it appears in $A_b(\mathbf{x})$ as a
bright, spatially extended ridge tracing the feature, while an isotropic
Gaussian field spreads power evenly across wedges and produces only diffuse,
uncorrelated speckle. As a result, $A_b(\mathbf{x})$ becomes an excellent diagnostic at visualizing \emph{where} the domain wall deposits the most power.

Before describing the two transforms in detail, we fix the numerical setup
common to both. All maps are simulated on a $512 \times 512$ Cartesian grid
subtending $10^\circ \times 10^\circ$, so that a single pixel subtends
$\theta_\mathrm{pix} = 1.17\,\mathrm{arcmin}$, comparable to the SPT-3G
$150\,\mathrm{GHz}$ beam. Primordial skies are drawn as Gaussian random fields
from a flat-sky $C(\ell)$ template with acoustic peaks, and the wall-lensed map
is obtained from the \emph{same} realization by the coordinate remapping of
Section \ref{sec:CMBDeflection}. White instrumental noise at
$10\,\mu\mathrm{K}\,\mathrm{arcmin}$, equivalently
$\sigma_\mathrm{pix} = 8.5\,\mu\mathrm{K}$ per pixel or about $8.5\%$ of the CMB
rms, is added independently to each map after lensing. Because the Gaussian
reference $T_{\mathrm{ref}}(\hat{n})$ and the lensed map
$T_{\mathrm{wall}}(\hat{n})$ share the same underlying primordial sky, any
difference between their coefficient statistics is attributable to the wall
deflection alone and not to cosmic variance between independent realizations.
The noise realizations, by contrast, are drawn independently, so the residual
scatter in the paired difference is set by the instrument noise rather than by
the primary CMB.

The wavelet analysis uses the orthogonal Daubechies-12 (\texttt{db12}) basis
\cite{Daubechies:1992}, a smooth, compactly supported wavelet with twelve
vanishing moments, applied as a separable two-dimensional discrete wavelet
transform. We decompose each map to four
levels. At each level $j$ the transform returns three detail sub-bands, the
horizontal $c_H^{(j)}$, vertical $c_V^{(j)}$ and diagonal $c_D^{(j)}$
coefficients, together with a residual approximation array that is discarded.
Level $j$ isolates structure with characteristic wavelength between $2^{j}$ and
$2^{j+1}$ pixels, so that the four levels probe
$2.3$--$4.7$, $4.7$--$9.4$, $9.4$--$18.8$ and $18.8$--$37.5\,\mathrm{arcmin}$,
corresponding to $\ell \sim 4600$--$9200$, $2300$--$4600$, $1150$--$2300$ and
$580$--$1150$ respectively. The three sub-bands are then pooled into a
single coefficient vector per level, containing $1.1\times 10^{5}$ values at
level $1$ down to $1.7 \times 10^{3}$ at level $4$.

\begin{figure}[h]
    \centering
    \includegraphics[width=1\linewidth]{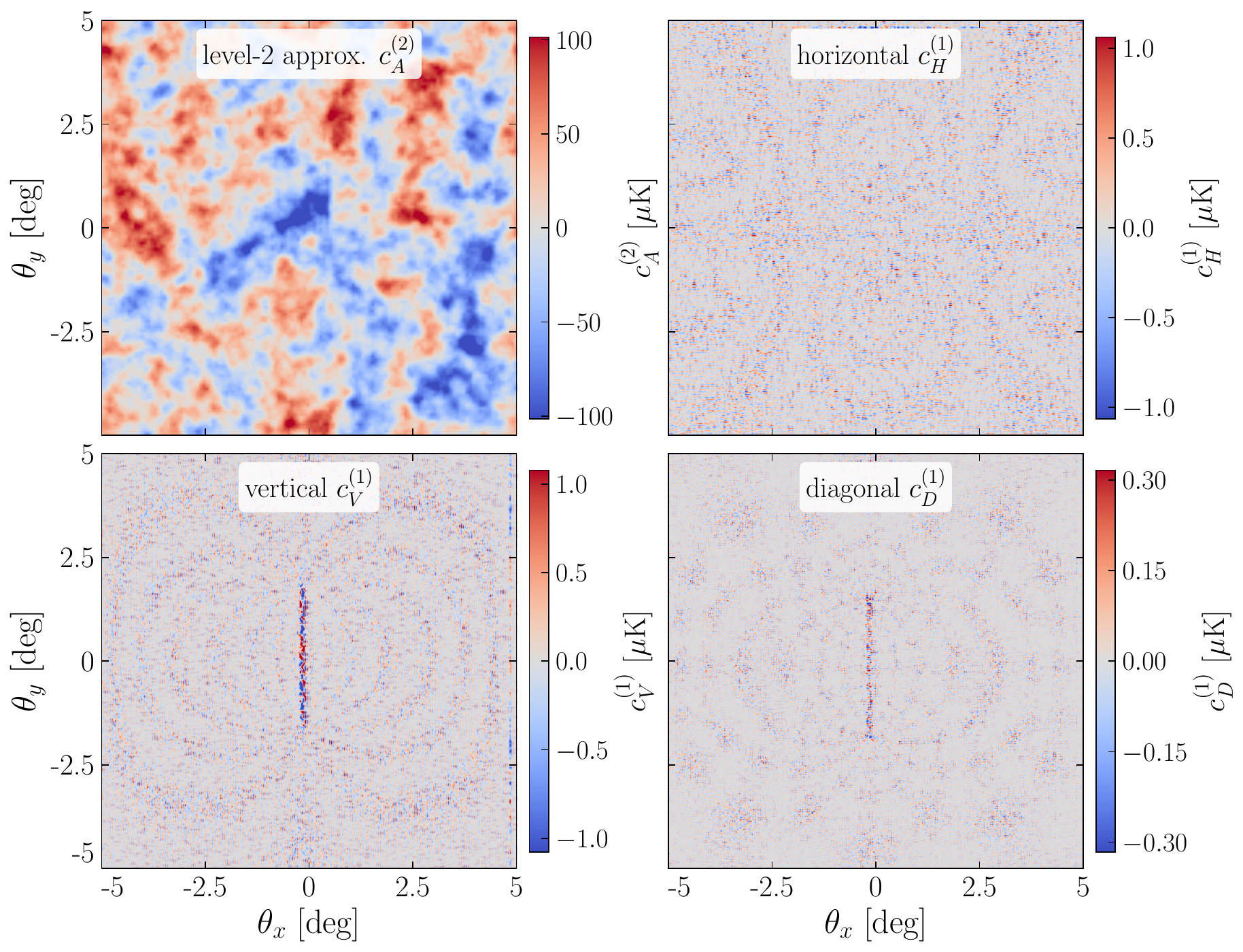}
    \caption{Wavelet detail sub-bands ($c_H^{(1)}$, $c_V^{(1)}$, $c_D^{(1)}$)
together with the level-2 approximation ($c_A^{(2)}$), obtained from a
Daubechies-12 (db12) decomposition of the domain-wall-lensed CMB temperature
field. It is observed that the coherent signature of the deflection is imprinted on the coefficients. $c_A^{(2)}$ is the level-2 approximation (low-pass) coefficient array of the db12 wavelet decomposition, i.e. the smooth, coarse-scale residual of the lensed CMB temperature map left after the finer level-1 and level-2 detail sub-bands have been stripped off, leaving behind the large angular scales where the CMB power is concentrated. }
    \label{fig:Wavelet}
\end{figure}

The separable construction gives this transform only a coarse three-way
directional selectivity: $c_H$, $c_V$ and $c_D$ respond preferentially to edges
near $0^\circ$, $90^\circ$ and $\pm 45^\circ$, with no finer resolution in
between, and pooling the three renders the statistic orientation-agnostic. The
wavelet transform therefore acts as a sensitive but essentially undirected
detector of localised non-Gaussianity: it registers the presence of an edge
without being tuned to its orientation. This makes it the natural baseline
against which the directional gain of the curvelet transform is measured. An
example wavelet amplitude map can be found in Figure \ref{fig:Wavelet}.

The curvelet analysis uses a directional multiscale decomposition constructed
directly in the Fourier domain, in the spirit of the discrete curvelet
transform of \cite{Starck:2002}. We compute the two-dimensional Fourier
transform $\tilde{T}(\mathbf{f})$ of the map and partition the
frequency plane in polar coordinates $(f, \varphi)$, with
$f = |\mathbf{f}|$ measured in cycles per pixel and
$\varphi = \mathrm{atan2}(f_y, f_x)$. A curvelet coefficient array is obtained
by multiplying $\tilde{T}$ by the product of a radial and an angular window and
inverse transforming,
\begin{equation}
    C_{b,k}(\mathbf{x}) \;=\; \mathscr{F}^{-1}\!\left[
      \tilde{T}(\mathbf{f})\,
      U\!\left(\frac{f-f_b}{\Delta f_b}\right)
      U\!\left(\frac{\Delta\varphi_k}{\Delta\varphi}\right)
    \right](\mathbf{x}),
    \label{eq:curvelet-def}
\end{equation}
where $\mathscr{F}^{-1}$ is the inverse Fourier transform, $b$ labels the radial band, $k$ the angular wedge,
$f_b$ and $\Delta f_b$ are the centre and half-width of band $b$, and
$\Delta \varphi_k = \arg\!\left[ e^{i(\varphi - \varphi_k)} \right]$ is the
signed angular separation from the wedge centre $\varphi_k$, wrapped to
$(-\pi, \pi]$. Both windows use the same raised-cosine profile
\begin{equation}
    U(u) =
    \begin{cases}
      1, & |u| < 1-\delta, \\[2pt]
      \tfrac{1}{2}\left[ 1 + \cos\!\left( \pi \dfrac{|u|-1+\delta}{\delta} \right) \right],
        & 1-\delta \le |u| \le 1, \\[6pt]
      0, & |u| > 1,
    \end{cases}
    \label{eq:curvelet-window}
\end{equation}
with taper fraction $\delta = 0.15$. The smooth roll-off suppresses the ringing
that a hard-edged frequency mask would introduce, at the cost of a controlled
overlap between neighbouring wedges. Prior to the forward transform each map is
multiplied by a separable Tukey window \cite{Tukey1977} of taper parameter $0.1$, which removes
the spectral leakage that the discontinuity at the patch boundary would
otherwise scatter across all wedges.

We use two radial bands, $f \in [0.10, 0.25]$ and $f \in [0.25, 0.45]$ cycles
per pixel. On our $1.17\,\mathrm{arcmin}$ grid these correspond to wavelengths
of $4.7$--$11.7\,\mathrm{arcmin}$ ($\ell \sim 1840$--$4600$) and
$2.6$--$4.7\,\mathrm{arcmin}$ ($\ell \sim 4600$--$8300$). We refer to the
latter as the finest band $B_2$, which carries the bulk of the
wall signal, since the deflection produces its sharpest gradients on the scale
of the void boundary. The lower cutoff $f = 0.10$ excludes the largest modes,
which are dominated by the primordial acoustic peaks and are poorly sampled on
a $10^\circ$ patch, while the upper cutoff $f = 0.45$ sits just inside the
pixel Nyquist frequency.

Each band is tiled with $2 n_\varphi = 16$ angular wedges, obtained by placing
the centres at $\varphi_k = k\pi/n_\varphi - \pi$ for
$k = 0, \dots, 2n_\varphi-1$ with $n_\varphi = 8$, so that the wedge centres
are separated by $\Delta\varphi = \pi/n_\varphi = 22.5^\circ$ and span the full
circle. Each wedge has angular support $|\Delta\varphi_k| < 22.5^\circ$, that is
a full width of $45^\circ$, so that adjacent wedges overlap by a factor of two
and every frequency is assigned to at most two wedges. The tiling is therefore
a redundant, non-orthogonal frame rather than a partition of unity. This is
adequate for our purpose, since we use the decomposition purely as an analysis
filter bank and never require reconstruction of the map from its coefficients.
We note that because $T$ is real, $\tilde{T}$ is Hermitian and the wedges at
$\varphi_k$ and $\varphi_k + \pi$ return complex-conjugate coefficient arrays,
$C_{b,k+n_\varphi} = C_{b,k}^{*}$. The $16$ wedges thus sample $8$ independent
orientations at $22.5^\circ$ spacing, each represented twice. The redundancy
merely symmetrizes the imaginary part of the pooled coefficient distribution and
adds no information. As a result, in Eq.~\eqref{eq:curvelet-amplitude} it renders
$A_b(\mathbf{x})$ uniformly a factor $\sqrt{2}$ larger than the
eight-wedge quantity, which is unimportant since the amplitude maps are displayed
on a shared, arbitrary colour scale.

Unlike the separable wavelet, each curvelet element is elongated in real space,
with an aspect ratio set by the ratio of the angular to the radial bandwidth,
and is tuned to a specific orientation. A coherent line-like feature therefore
deposits its power into the handful of wedges whose normal is aligned with the
feature, leaving the remainder near zero, whereas an isotropic Gaussian field
spreads power evenly across all $8$ orientations. This anisotropic,
edge-matching response is precisely what makes curvelets effective at detecting
the planar geometry of a domain wall.

\begin{figure}[t]
    \centering
\includegraphics[width=1\columnwidth]{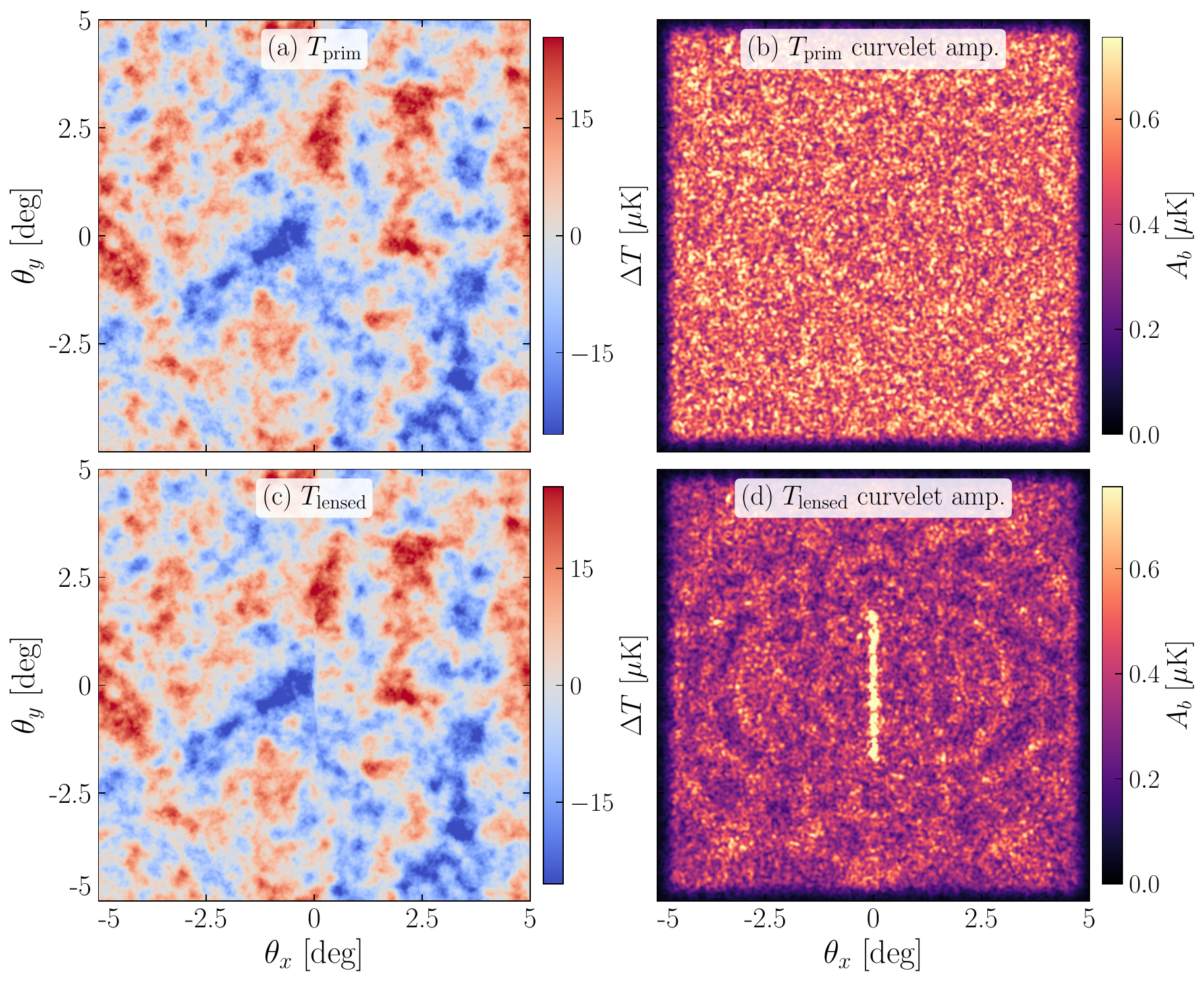}
    \caption{Finest-band curvelet amplitude $A_b(\mathbf{x})$
        (Eq.~\ref{eq:curvelet-amplitude}, band $B_2$) for a Gaussian reference map (top)
        and the same sky lensed by a domain wall (bottom). Left: input CMB
        temperature maps; right: orientation-summed curvelet amplitudes on a
        shared colour scale. The wall produces a bright coherent ridge in \emph{d)}, absent in the Gaussian case \emph{b)}, while the lensing is close to invisible by eye in \emph{c)}.
        }
    \label{fig:curvelet}
\end{figure}

Figure \ref{fig:curvelet} shows this comparison, evaluated on the finest band
$B_2$. The left column displays the
input temperature maps and the right column the corresponding curvelet
amplitudes, with the two amplitude panels sharing a common colour scale
so that edge brightness can be compared directly. For the Gaussian reference,
the amplitude map is featureless: power is distributed roughly uniformly across
the patch, consistent with a statistically isotropic field with no preferred
direction. For the wall-lensed map, by contrast, $A_b(\mathbf{x})$ deposits a sharp
coherent ridge of curvelet amplitude along the wall, marking the
location where the deflection field produces a discontinuity in the temperature
gradient.

While this is an excellent visualization, for our detection statistic we do not
reduce to the amplitude map. Pooling $|C_{b,k}(\mathbf{x})|$ in quadrature discards the
sign of the coefficients, the very feature that we exploit. The wall imprints a
two-sided tail on the \emph{signed} coefficient distribution because the
deflection field is antisymmetric about the wall plane, producing gradient
excursions of both signs at the two void boundaries. Taking the modulus folds
these onto a single positive tail and, in doing so, converts a fourth-moment
excess into a shift of the mean, which is far less distinguishable from the
null. We therefore build the detection statistic from the excess kurtosis of the
signed curvelet coefficients, pooling the real and imaginary parts over all
wedges within each radial band. The wavelet statistic is constructed
analogously, from the excess kurtosis of the signed detail coefficients
$(c_H, c_V, c_D)$, computed separately at each of the four db12 decomposition
levels. This kurtosis can be computed as
\begin{equation}
  \mathcal{K}_b=\frac{\tfrac1N\sum\limits_n (x_n-\bar x)^4}
                {\big[\tfrac1N\sum\limits_n(x_n-\bar x)^2\big]^2}-3,
  \label{eq:kurtosis}
\end{equation}
where $x$ is the pooled vector of signed coefficients of channel $b$, $\bar{x}$
is its sample mean, and the $-3$ term fixes $\mathcal{K}_b = 0$ for a Gaussian.

We refer to each of these coefficient sets as a \emph{channel} $b$. There are
six in total: four wavelet levels and two curvelet radial bands. For the
wavelet channels, $x$ is the concatenation of the
$(c_H^{(j)}, c_V^{(j)}, c_D^{(j)})$ coefficients at level $j$. For the curvelet
channels, $x$ is the concatenation of the real and imaginary parts of
$C_{b,k}(\mathbf{x})$ over all wedges $k$ and all pixels $\mathbf{x}$, giving
$8.4 \times 10^{6}$ pooled values per band. Pooling across wedges before taking
the moment is deliberate: the projected orientation of a wall is not known a
priori, and a statistic built from a single preselected wedge would depend on
that unknown. The kurtosis of the pooled distribution is elevated whenever
\emph{some} orientation carries anomalous weight, without our having to
identify which. For an approximately Gaussian primordial sky both wavelet and
curvelet kurtoses sit near zero, while a wall-lensed map drives them to strongly
elevated values, and it is this contrast that the following section quantifies
through the discriminability statistic $d'$.

\section{Simulation-Based Inference\label{sec:SBI}}

The lensing signature of a domain wall is spatially localised, anisotropic, and morphology dependent. Its statistical properties are non-Gaussian and sensitive to combinations of physical parameters. Standard likelihood based methods, which typically rely on Gaussian approximations and compression into angular power spectra, are poorly suited to this signal: the relevant information is encoded in the spatial structure of the deflection pattern rather than in its two-point statistics. Simulation-based inference (SBI) provides a natural alternative \cite{Cranmer:2020,TejeroCantero:2020}, bypassing the need for an explicit likelihood function entirely and instead making inferences directly from samples of forward simulations. Neural network-based SBI has already been applied to defect searches in CMB temperature maps \cite{Ciuca:2019, Ciuca:2019b,VafaeiSadr:2018,Torki:2022}.
In the following, we will demonstrate the effectiveness of SBI in this problem via Neural Ratio Estimation (NRE) for setting upper limits in the weak-signal regime, and Neural Posterior Estimation (NPE) for characterization if a signal is observed.

\subsection{Neural Posterior Estimation}

Before introducing the details of our pipeline, we summarize here the main idea behind NPE. Without an explicit likelihood, the starting point is a simulator, or forward model, $\mathcal{F}$ that maps values in the parameter space $\Theta$ to a mock observation $x=\mathcal{F}(\Theta,\nu)$. In our case, the data space is the 2-D temperature map $\mathbb R^{256\times256}$. The forward model is stochastic: each run also draws a set of \emph{nuisance parameters} $\nu$, which for us describe both the unlensed CMB temperature realization and the detector noise. The likelihood of data $x$ given $\Theta$ is the integral
\begin{equation}
  p(x \mid \Theta) = \int \! \dd\nu \; p(\nu) \;
     \delta\big(\,x - \mathcal{F}(\Theta,\nu)\big).
  \label{eq:npe-likelihood}
\end{equation}
Since $\nu$ carries a dimension comparable to that of the data itself, this integral cannot be realistically evaluated, which excludes the possibility of an explicit likelihood analysis. What the simulator gives us, at comparably low cost, are exact samples from the joint distribution $p(\Theta,x) = p(\Theta)\,p(x\mid\Theta)$, and as we will show, this is sufficient for parameter inference. We start by describing the mathematical structure of the posterior estimator.

NPE replaces the intractable posterior with a parametrized family of probability densities on the parameter space that depends on the observed data, $q_\varphi(\Theta \mid x)$. The densities
each factorizes into a compression stage $f_\psi$ and a flow stage $p_\phi$ as follows
\begin{equation} \label{eq:npe-family}
\begin{aligned}
    c &= f_\psi(x)\\
    q_\varphi(\Theta \mid x) \;&\equiv\;
    p_\phi\!\left(\Theta \mid c\right)
\end{aligned}
\end{equation}
Here $f_\psi$ is the \emph{featurizer} that takes the data $x$ to a \emph{context vector} $c$, a learned summary of the data with significantly lower dimensionality than the data. This first stage is realized as a Convolutional Neural Network (CNN) which repeatedly convolves the data with learned kernels to produce $c$. The learned parameters of the CNN is denoted as $\psi$. For details of our CNN architecture, please see Section \ref{sec:NPEArch}.

The second stage $p_\phi(\,\cdot \mid c)$ is a probability density on the $\Theta$ parameter space that depends on the context vector $c$, and hence informed by data. It is realized as a \emph{normalizing flow}, i.e. a parametrized family of diffeomorphisms $g_\phi(\,\cdot\,;c)$ mapping variables in a \emph{latent space} $u\in\mathbb R^{d}$ to the parameter space $\Theta\in\mathbb R^{d}$. The posterior is the pushforward of a simple distribution on the latent space, usually a unit Gaussian, under $g_\phi$. The probability density can be obtained by transforming with the Jacobian,
\begin{equation}
  p_\phi(\Theta \mid c) =
    \mathcal{N}\!\left(u;\,\mathbf{0},\mathbbm{1}_d\right)
    \left| \det \frac{\partial u}
                     {\partial \Theta} \right| ,
  \label{eq:npe-flow}
\end{equation}
with $u=g_\phi^{-1}(\Theta;c)$ and $\mathcal{N}(u;\mathbf{0},\mathbbm{1}_d)$ the unit Gaussian in the latent space. Since $g_\phi$ is invertible, the network evaluates the posterior at any $\Theta$ by evaluating $g_\phi^{-1}$, and samples
from it by drawing a Gaussian point and evaluating $g_\phi$. We will detail our implementation of $g_\phi$ later in Section \ref{sec:NPEArch}.

Finally, $q_\varphi$ is simply the composition of $p_\phi$ and $f_\psi$, and $\varphi$ denotes all the trainable parameters, the concatenation $\varphi=(\phi,\psi)$. The two stages are trained together, $\phi$ and $\psi$ varying simultaneously, by minimizing the negative log-density the model assigns to the parameters that actually produced each map,
\begin{eqnarray}
  \mathcal{L}(\varphi) &=& - \, \mathbb{E}_{(\Theta,x) \sim p(\Theta,x)}
    \!\left[ \ln q_\varphi(\Theta \mid x)\right]
    \nonumber \\
  &\simeq& -\frac{1}{N}\sum_{i=1}^{N}
    \ln q_\varphi(\Theta_i \mid x_i),
  \label{eq:npe-loss}
\end{eqnarray}
where $\mathbb{E}_{(\Theta,x)\sim p(\Theta,x)}$ is the average over parameter--map pairs produced by the simulator and the second line is its Monte-Carlo estimate over the $N$ simulated pairs $(\Theta_i,x_i)$. That minimizing $\mathcal{L}$ returns the posterior follows from writing $p(\Theta,x) = p(x)\,p(\Theta\mid x)$ and inserting the exact posterior inside the logarithm,
\begin{equation}
  \mathcal{L}(\varphi) = \mathbb{E}_{x \sim p(x)}
    \Big[ D_\mathrm{KL}\!\left( p(\Theta \mid x) \,\big\|\,
      q_\varphi(\Theta \mid x) \right) \Big]
    \; + \; \mathcal{H}\!\left[\Theta \mid x\right],
  \label{eq:npe-kl}
\end{equation}
where $D_\mathrm{KL}(p \,\|\, q) = \int \dd\Theta \; p \ln(p/q)$ is the Kullback--Leibler divergence, a measure of distance between two distributions, and
$\mathcal{H}[\Theta\mid x] = -\mathbb{E}_{p(\Theta,x)}[\ln p(\Theta\mid x)]$ is the conditional entropy of the exact posterior, a fixed number set by the physics of the problem and independent of $\varphi$. The divergence is non-negative and vanishes only when its two arguments coincide in the support of $p(\Theta \mid x)$. Training is thus a variational problem whose unique optimum is the true posterior. Finally, the estimator is \emph{amortised}: $\varphi$ is fit once, after which the posterior for any new map costs a single forward pass.

The following caveats apply: the optimum is only well approximated if the family (\ref{eq:npe-family}) is flexible enough to contain the true posterior, the training set large enough to resolve it, and the compression is lossless enough in that $c$ is a sufficient statistic for $\Theta$, since whatever information $f_\psi$ discards raises the floor in loss (\ref{eq:npe-kl}). The performance however can be checked empirically by a calibration test, which we carry out in Figure \ref{fig:PPplot}.
We stress that, because $\psi$ is optimized under the same loss as $\phi$, the summary $c$ is learned rather than imposed by hand, in contrast to the power spectrum or to the curvelet kurtosis of Section \ref{sec:Curvelet}: the compression is tuned to retain exactly what bears on $\Theta$.

\subsubsection{Training data}

We here describe in details the NPE pipeline specific to our problem, starting from the training data generation. The physical parameters of interest are the scattering redshift $z$, energy scale $\eta$ (related to the surface density via $\sigma \propto \eta^3$), and inclination angle $\theta$ with respect to the line of sight. For all simulations, the clustering due to domain walls begins at matter-radiation equality, $z \approx 3400$. These are sampled from the prior distributions
\begin{eqnarray}
    z &\sim& \mathcal{U}(5,\,100), \nonumber \\
    \eta &\sim& \mathcal{U}(1\,\text{MeV},\,100\,\text{MeV}) \nonumber \\
    \theta &\sim& \mathcal{U}(1^\circ,\,89^\circ) 
\end{eqnarray}
and $2\%$ of maps are instead reassigned $\eta = 0$ so that the network samples the no-wall-sky too. The energy scale is sampled uniformly in $\eta$ rather than in $\sigma$ to ensure linear coverage of the $(\eta/\text{MeV})^3$ prior, avoiding the overrepresentation of high-tension walls that would result from uniform sampling in $\sigma$ directly.

For each parameter realisation, an unlensed CMB temperature map is generated as a Gaussian random field whose power spectrum is chosen to reproduce the acoustic peak structure of the observed CMB \cite{Planck:2018V,Planck:2018}. The two-component angular deflection field $(\alpha_x, \alpha_y)$ is then computed analytically for the given wall parameters using the relativistic domain wall deflection formalism described in Section \ref{sec:Deflection}, and the lensed map is obtained via
\begin{equation}
  T_\mathrm{obs}(\hat{n}) = T_\mathrm{prim}\!\left(\hat{n}
    - \boldsymbol{\alpha}(\hat{n})\right),
\end{equation}
where the remapping is performed using bilinear interpolation on a regular angular grid. Each $256 \times 256$ pixel map covers a $10^\circ \times 10^\circ$ field of view. A total of 600,000 lensed maps were generated, split into 550{,}000 training and 50{,}000 validation examples. White detector noise is added in pixel space at a map depth of $10\,\mu\mathrm{K}\,\mathrm{arcmin}$, matching the depth adopted for the wavelet and curvelet analyses of Section \ref{sec:Curvelet}; on the $\theta_\mathrm{pix} = 2.34\,\mathrm{arcmin}$ grid used here this corresponds to $\sigma_\mathrm{pix} = 4.3\,\mu\mathrm{K}$ per pixel, or roughly $4\%$ of the CMB rms. Together with the primordial sky realization, this noise realization constitutes the nuisance parameters $\nu$ of (\ref{eq:npe-likelihood}), and it is redrawn independently each time a map is shown during training, so that no noise realization is seen twice and the estimator cannot key on a fixed noise pattern. The same noise treatment is applied to the maps used for sensitivity estimation in Section \ref{sec:NreConst}. The $2.34\,\mathrm{arcmin}$ pixelization of the SBI maps is already coarser than the SPT-3G $150\,\mathrm{GHz}$ beam at $1.4 \,\mathrm{arcmin}$, so it is the pixel scale, and not the beam, that sets the smallest resolved angular scale in our simulations. We utilize the coarser resolution due to the computational cost of generating a large data set of lensed maps, quadrupling the cost at the SPT-resolution. The noise is taken to be white and homogeneous, with no atmospheric $1/f$ component and no large-scale filtering, as in Section \ref{sec:Curvelet}. This idealized model is utilized to focus on the wall signal, while checking that the network can learn the signal at all above the noise floor.

The ratio estimator of Section \ref{sec:NRE} is trained on a separate set of maps, generated with a refined forward model. The primordial skies are Gaussian random fields with a phenomenological power spectrum fitted to the acoustic peaks and damping tail, normalized to an rms of $100\,\mu\mathrm{K}$. Each is generated on a $35^\circ \times 35^\circ$ source plane from which the lensed $10^\circ \times 10^\circ$, $256 \times 256$ pixel patch is resampled, so that the source position of every pixel lies inside the simulated sky for all walls in the prior. Each source map is generated in Fourier space with no power above the Nyquist frequency of its grid, so by the Nyquist--Shannon sampling theorem \cite{Nyquist:1928,Shannon:1949} the field between pixels is fixed exactly by the pixel values. We use this to upsample each source map by a factor of two in resolution, zero-padding its Fourier transform so that the inverse transform returns the same band-limited field on a grid with half the pixel spacing. A cubic spline on this finer grid is then evaluated at the displaced positions $\hat{n} - \boldsymbol{\alpha}$. This is done because the original grid was too sparse for cubic spline interpolation. The scheme preserves the power at $\ell = 2400$--$4000$ to $0.3\%$ for a pure translation, compared with a lensing imprint of $0.1$--$2.5\%$ in that band, whereas simpler interpolation on the native grid could lose several percent. Null maps pass through the same resampling. The maps are then convolved with a Gaussian beam of $1.4\,\mathrm{arcmin}$ FWHM in harmonic space, and white noise at $10\,\mu\mathrm{K}\,\mathrm{arcmin}$ is added, redrawn each time a training map is loaded.

\subsubsection{NPE Architecture}
\label{sec:NPEArch}

We now specify the two stages of NPE described in (\ref{eq:npe-family}) concretely, with $f_\psi : \mathbb{R}^{256\times 256} \to \mathbb{R}^{128}$ and $d=3$. The featurizer is a CNN that progressively coarse-grains the $256 \times 256$ map in four downsampling stages, each halving the spatial resolution and doubling the number of feature channels, so that the network builds up increasingly large-scale, abstract features of the map. A global pooling step then collapses this into a fixed-length vector, which is passed through a small Multi-Layer Perceptron (MLP) to produce the 128-dimensional context vector. The featurizer opens with a $5\times5$ convolution to $32$ channels, followed by four stages, each made of two residual blocks (group normalization, SiLU activation \cite{elfwing2018sigmoid} and a $3\times3$ convolution. The entire three-step sequence is applied twice, with a skip connection) and a stride-$2$, $3\times3$ convolution that halves the resolution and doubles the number of channels, from $32$ to $512$. A global average over the remaining $16\times16$ grid is passed through a two-layer MLP ($512\to256\to128$) to give the context vector. The featurizer contains approximately 4.9 million trainable parameters.

The normalising flow is a six-layer masked autoregressive flow (MAF)
\citep{papamakarios2017masked} conditioned on the context vector. The \emph{autoregressive} construction orders the three parameters and lets each depend only on those preceding it, so that the Jacobian in (\ref{eq:npe-flow}) is triangular and its determinant reduces to a product of $d$ diagonal entries; stacked layers then simply add their log-determinants. Stacking several layers gives the flow the flexibility to represent the non-Gaussian, correlated posteriors expected for this problem.

The full network, with a total of 5.6 million parameters, is trained jointly with the featurizer by minimizing the negative log-likelihood of (\ref{eq:npe-loss}).
We adopt the AdamW optimizer \cite{Loshchilov:2019} with an initial learning rate of $3 \times 10^{-4}$, reduced on a validation-loss plateau, with gradient clipping at unit norm for numerical stability. The $6 \times 10^5$ maps are streamed from disk in randomized batches of $10^4$ so that the network sees the full training set roughly $5$ times over the course of training. The loss-epoch curves and posterior calibration plot show that the training has sufficiently converged (Appendix \ref{sec:AppCalibration}).

\subsubsection{NPE Results}

\begin{figure}
    \centering
\includegraphics[width=\linewidth]{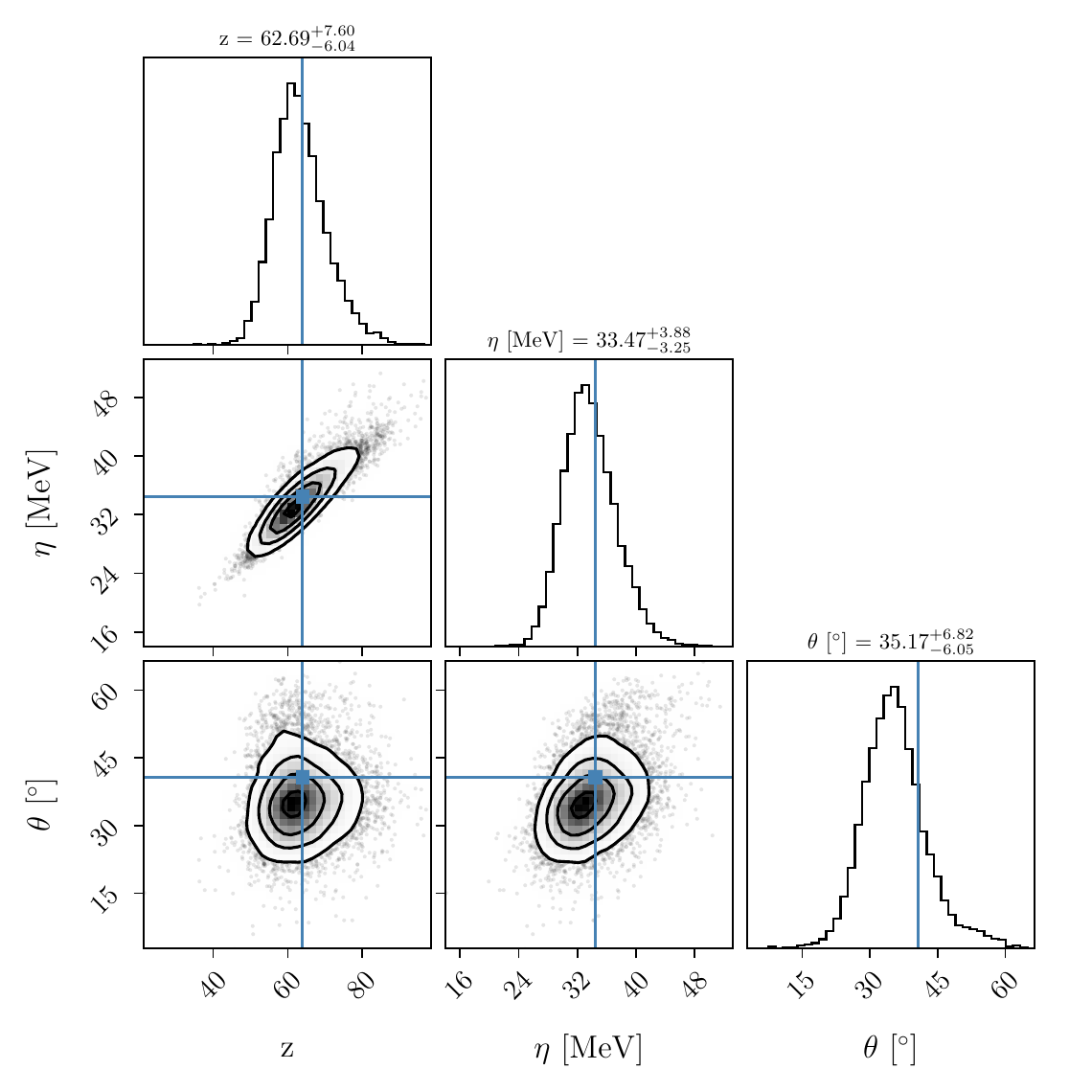}
        \caption{Representative posterior corner plot for a single validation observation, showing marginal and joint posteriors over $z$, $\eta$, and $\theta$ with the true parameter values marked. The elongated $z$--$\eta$ contours trace the degeneracy between a stronger wall at higher redshift and a weaker wall at lower redshift, while $\theta$ is uncorrelated with both.}
    \label{fig:Cornerplot}
\end{figure}

The inferred posteriors, shown in Figure \ref{fig:Cornerplot}, reflect the physical structure of the problem. The
redshift $z$ and energy scale $\eta$ exhibit a pronounced positive correlation in
the joint posterior. As discussed in Section \ref{sec:Deflection}, raising $\eta$
and lowering $z$ both strengthen the deflection field, so a stronger wall at
higher redshift produces a deflection amplitude similar to that of a weaker wall
at lower redshift. The degeneracy is only partial, since the location of the
overdense shell is set by the horizon $r_H(z)$ and not by $\eta$, which closes
the contours along the degeneracy direction. The inclination $\theta$, by
contrast, shows no appreciable correlation with either $z$ or $\eta$,
consistent with it being fixed by the projected morphology of the deflection
pattern rather than by its amplitude.

\subsection{Neural Ratio Estimation\label{sec:NRE}}

The neural posterior estimator of the preceding section learns the full
posterior $p(\Theta \mid x)$ and is well suited to characterising the joint
parameter structure of a detected wall. For the purpose of setting limits,
however, the likelihood ratio is a much more useful tool, and we therefore complement the NPE with a Neural Ratio Estimator \cite{Hermans:2020,Cranmer:2020}. Whereas the NPE models a density over the parameter space, the NRE
learns the likelihood-to-evidence ratio
\begin{equation}
  r(x,\,\Theta) = \frac{p(x \mid \Theta)}{p(x)}
\end{equation}
directly, by training a binary classifier to distinguish ``matched'' pairs
$(x,\Theta)$ drawn from the joint distribution $p(x,\Theta)$ from
``scrambled'' pairs $(x,\Theta')$ in which the parameters are drawn
independently of the map from the marginal prior. The optimal classifier
$d(x,\Theta)$ satisfies $d/(1-d) = r(x,\Theta)$, so the log-odds output of the trained
classifier, $\ln[d/(1-d)]$, is precisely $\ln r$; no explicit density estimator is required \cite{Hermans:2020}. The NRE uses the residual convolutional featurizer of Section \ref{sec:NPEArch} with identical
architecture. However, the two networks are trained independently, on separate training sets, and share no weights. The normalizing flow is replaced by a ratio head, so that the network returns $\ln r$ directly: the parameters pass through a two-layer encoder of width $64$, are concatenated with the $128$-dimensional context vector, and are mapped through three hidden layers of width $256$ to the scalar $\ln r$. The network is trained by minimizing the binary cross-entropy, the standard classification objective whose minimum is reached when the classifier outputs the true probability that a pair is matched. Scrambled pairs are formed within each minibatch of $48$ maps by cyclically shifting the parameter rows by a random non-zero offset, so that the two classes are equal in size and no pair is scrambled onto itself. We use AdamW with an initial learning rate of $3\times10^{-4}$, halved on each plateau of the validation loss, and adopt the checkpoint of minimum validation loss.

We construct the ratio over the two parameters of interest for the sensitivity
forecast, $(z,\eta)$, while the inclination $\theta=0^\circ$ is fixed to the edge-on wall to determine the idealized sensitivity. We make use of $40,000$ unique domain walls each drawn from the following distributions
\begin{eqnarray*}
    z &\sim& \mathcal{U}(5, \ 100) \qquad \theta = 0^\circ \\ \eta &\sim& 0.9\,\mathcal{U}(1\rm MeV, \ 100\rm MeV ) + 0.1\,\delta(\eta)
\end{eqnarray*}
where each wall is made to lens $4$ independent CMB realizations, for a total of $160,000$ lensed maps. To ensure that the network sufficiently trains on null skies, the second term assigns $10\%$ of the walls exactly $\eta=0$, the no wall limit, so that the null hypothesis lies inside the training support. Of these maps, $144{,}600$ are used for training and a fixed subsample of $4096$ of the remaining $15{,}400$ for validation. Every wall is centred in the field with a fixed orientation, so the forecast applies to a wall of known position and orientation and carries no look-elsewhere penalty over its placement.

The resulting $r(x,z,\eta)$ furnishes, for any observed map, a
two-dimensional likelihood-ratio surface. Working at fixed $z$ and treating
$\eta$ as the parameter to be constrained, we form the test statistic
\begin{equation}
  t(x;\,z,\eta) = 2\!\left[\ln r(x;\,z,\hat\eta) - \ln r(x;\,z,\eta)\right]
  \label{eq:teststat}
\end{equation}
where $\hat\eta$ maximizes $\ln r$ at that redshift. Since an upper limit excludes only values above the best fit, we use the one-sided statistic
\begin{equation}
  q(x;\,z,\eta) =
  \begin{cases}
    t(x;\,z,\eta), & \eta \ge \hat\eta,\\
    0, & \eta < \hat\eta.
  \end{cases}
  \label{eq:qstat}
\end{equation} For a wall of tension $\eta$, $q$ asymptotically follows $\tfrac{1}{2}\chi^2_0 + \tfrac{1}{2}\chi^2_1$ \cite{Chernoff:1954,Cowan:2011}, where $\chi^2_0$ is a point mass at zero, so that the probability density of $q$ is $\tfrac{1}{2}\delta(q) + \tfrac{1}{2}\,e^{-q/2}/\sqrt{2\pi q}$, the second term being the $\chi^2_1$ density: half of the time the best fit lies above the tested value and $q$ vanishes. The $95$th percentile of this mixture is the $90$th percentile of $\chi^2_1$, $t_{\rm crit} = 2.71$, so the $95\%$ upper limit lies where $\ln r$ has fallen $\Delta\ln r = 1.35$ below its maximum at that redshift. The familiar $\Delta\ln r = 1.92$ is the two-sided value and would give a $97.5\%$ upper limit. Because $\ln r$ is learned rather than exact, we check the asymptotic threshold on simulations. On the $8000$ null skies of Section \ref{sec:NreConst}, the statistic for excluding $\eta = 0$ exceeds $2.71$ in $4.4\%$ of skies, against the asymptotic $5\%$, and on validation maps containing walls the sampling distribution of $t$ follows $\chi^2_1$ away from the prior boundaries, and the resulting upper limit covers the true $\eta$ in $97.3\%$ of cases (Appendix \ref{sec:AppNRECalibration}). The threshold is therefore, if anything, mildly conservative.

\begin{figure}
    \centering
\includegraphics[width=\linewidth]{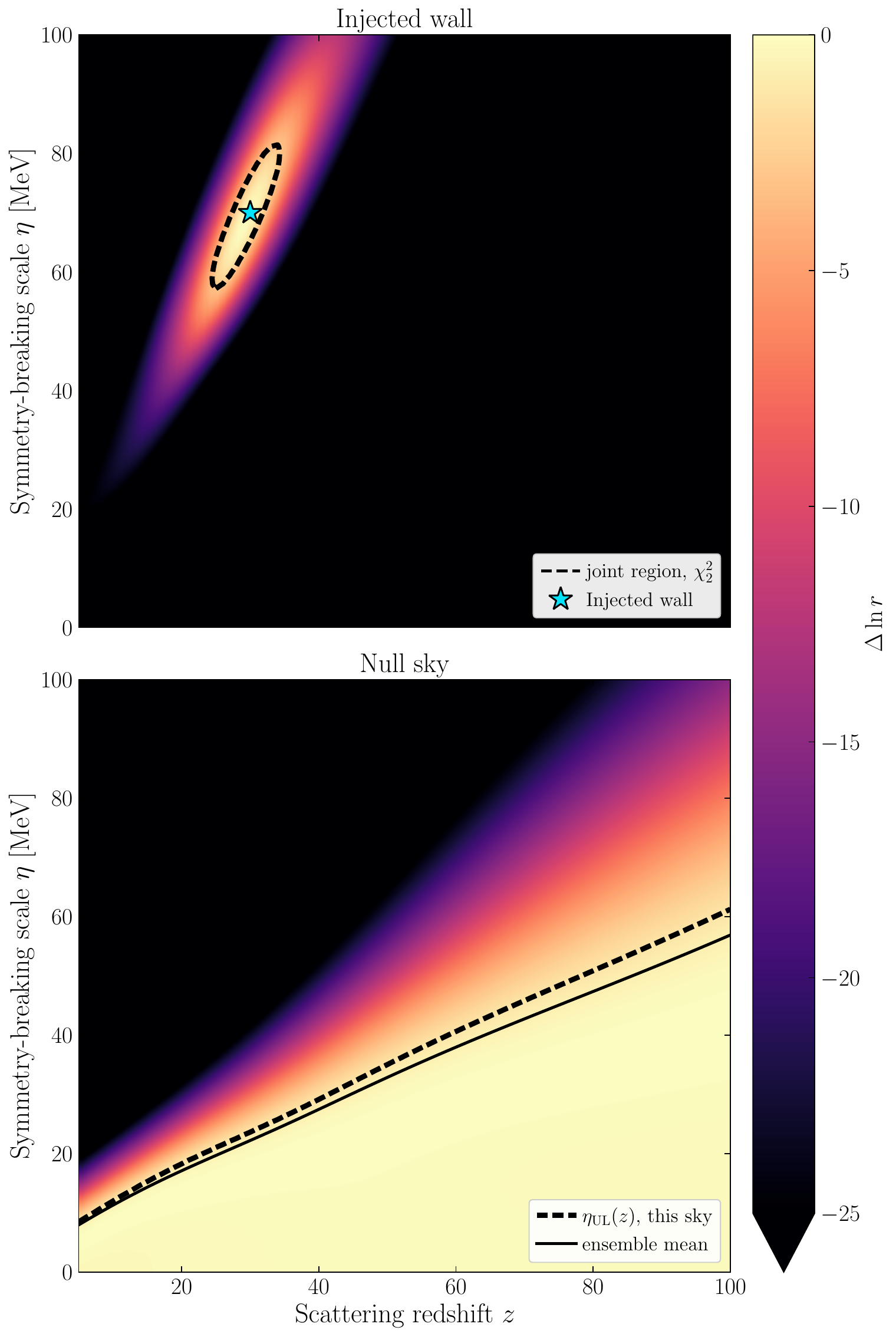}
    \caption{Learned log-ratio surface over the $(z,\eta)$ plane. \emph{Top:} a map lensed by an injected wall (star) at $(z_{\rm inj},\eta_{\rm inj}) = (30,\,70\,\mathrm{MeV})$, shown as $\ln r - \ln r_{\max}$. The surface peaks near the true parameters and is elongated along the $z$--$\eta$ degeneracy direction, confirming that the network has learned a physically meaningful ratio; the dashed contour is the joint $95\%$ confidence region, at $\Delta\ln r = 3.00$ for $\chi^2_2$. \emph{Bottom:} a sky without a wall, shown relative to its maximum at each redshift, $\ln r - \mathrm{max}_\eta \ln r$. Its dashed contour at $\Delta\ln r = 1.35$ ($t_{\rm crit} = 2.71$) is the conditional upper limit $\eta_{\rm UL}(z)$ this sky sets, and the filled curve is the mean of $\eta_{\rm UL}(z)$ over $8000$ null skies, the expected limit of Figure \ref{fig:Constraints}. The colour scale is clipped at $\Delta\ln r = -25$ to highlight the structure near the maximum.}
    \label{fig:NREInject}
\end{figure}

As a validation of the learned ratio, we evaluate $\ln r(x,z,\eta)$ on a map
lensed by an injected wall (Figure \ref{fig:NREInject}, top) and confirm that the surface peaks in the
neighbourhood of the true parameters, extended along the same $z$--$\eta$
degeneracy direction seen in the NPE posteriors Figure \ref{fig:Cornerplot}. The same degeneracy that limits
the NPE constraint on $\eta$ is therefore present here; the advantage of the NRE
is not that it breaks the degeneracy but that it yields a calibrated test
statistic for limit setting without requiring a well-localised posterior. The bottom panel shows the same surface for a sky without a wall, conditioned on the redshift, together with the upper-limit curve it yields.

\section{\label{sec:Constraints}Observational Constraints}
In this section, we look into 3 methods to probe domain walls. First, we look at an instrumental cutoff using the South Pole Telescope's (SPT) angular resolution and the deflection strength. Secondly, we make use of the kurtosis analysis of Section \ref{sec:Curvelet} to gauge the reach of statistical methods on an SPT-like observational grid. Thirdly, we make use of the SBI-NRE architecture developed in Section \ref{sec:SBI} to draw out a forecast sensitivity curve.

The curves derived below are all idealized sensitivity forecasts. They delineate the region of the $(z,\eta)$ plane in which a domain wall of a given tension and redshift would produce an imprint accessible to a given method, evaluated for a single, optimally aligned wall imprinted on an otherwise Gaussian sky. They therefore mark the idealized reach of each method, that is, where lensing \emph{can} be used to find a domain wall with statistical significance, and not a claim of exclusion of real walls by non-detection. A realistic search must additionally contend with $\Lambda$CDM lensing, foregrounds, point sources, inhomogeneous noise and beam systematics, none of which enter our maps. We adopt the South Pole Telescope \cite{SPT3G:2021} as a representative arcminute-resolution ground-based experiment, with a beam $\theta_\mathrm{beam} \approx 1.4\,\mathrm{arcmin}$ at $150\,\mathrm{GHz}$ and map noise level of $\mathcal{O}(10)\,\mu\mathrm{K}\,\mathrm{arcmin}$. Because the deflection amplitude depends on wall inclination, all amplitude-based curves are evaluated at the aligned value of $\theta = 0^\circ$.
\begin{figure*}
    \centering
    \includegraphics[width=\linewidth]{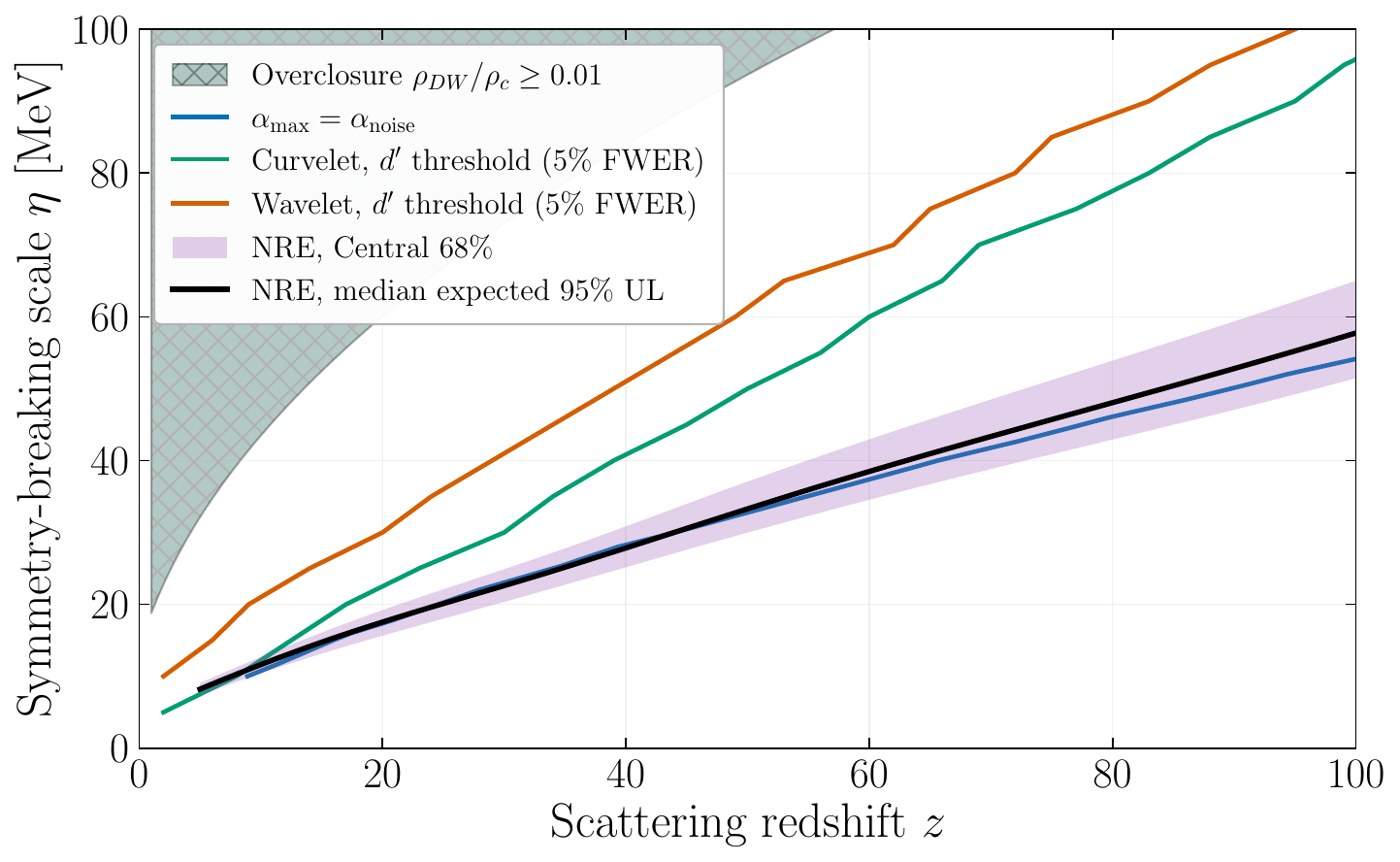}
    \caption{Forecasted sensitivity to cosmological domain walls in the $(z,\eta)$
plane, where $z$ is the redshift at which the observed photons scatter off the
wall and $\eta$ is the symmetry-breaking scale, related to the wall tension by
$\sigma \propto \eta^3$. Every curve is an idealized sensitivity forecast rather
than an exclusion. Each is computed for a single wall of finite extent $L(z)$,
aligned with the line of sight at $\theta = 0^\circ$ and centered on an
otherwise Gaussian $10^\circ \times 10^\circ$ patch, with white instrumental
noise at $10\,\mu\mathrm{K}\,\mathrm{arcmin}$ and no $\Lambda$CDM lensing,
foregrounds, point sources or beam systematics. Within the region \emph{above}
each curve the corresponding method registers such a wall at the stated
significance, so the curves mark the reach of each method and not a claim that
real walls in that region are excluded. Since the deflection amplitude falls
with inclination and vanishes for the face-on wall, the aligned orientation used
here is the most optimistic one, and a randomly oriented population would sit
above these curves. \textbf{Hatched region (Overclosure,
$\rho_{DW}/\rho_c \ge 0.01$):} the cosmological boundary at which the wall energy
density reaches $1\%$ of the critical density at redshift $z$, obtained from the
scaling solution $\rho_{\rm wall} \sim \sigma H$. This is the one genuine exclusion in the figure and
is independent of any instrument or statistic.
\textbf{Blue ($\alpha_\mathrm{max} = \alpha_\mathrm{noise}$):} the
per-resolution-element noise reference, at which the induced temperature shift
$\delta T = \boldsymbol{\alpha}\cdot\nabla T_\mathrm{CMB}$ equals the map noise
within a single beam. This is a per-pixel criterion and not an
information-theoretic floor. \textbf{Green
(Curvelet) and orange (Wavelet):} the kurtosis detection thresholds of
Section \ref{sec:KurtosisConst}, where the discriminability
$d' = \Delta\mathcal{K}/\sigma_\mathrm{null}$ of (\ref{eq:dprime})
crosses $1.96$ for the $M=2$ curvelet radial bands and $2.24$ for the $M=4$
wavelet levels. These follow from a Bonferroni-corrected one-sided per-channel
level $\alpha_1 = 0.05/M$ for a family-wise false-positive rate of $5\%$. Each
boundary is estimated from $100$ paired $10^\circ \times 10^\circ$ realizations
in which the wall-lensed sky and the Gaussian reference share the same
primordial realization, so that the pairing cancels cosmic variance. Because
$d'$ is normalized by the across-realization scatter of the null kurtosis rather
than by $\sigma_\mathrm{null}/\sqrt{N}$, it is an effect size rather than a
$t$-statistic, and the quoted significance is that attainable from a single
SPT-sized patch. The grid is scanned over $20$ tensions in steps of
$5\,\mathrm{MeV}$ and $120$ redshifts on $z \in [1,120]$, and is constrained to
be monotonically non-decreasing in $\eta$ as a deliberate regularization against
grid-level noise. The curvelet boundary is set almost everywhere by the finer
$B_2$ band spanning $2.6$--$4.7\,\mathrm{arcmin}$, and the wavelet boundary by
level $2$ spanning $4.7$--$9.4\,\mathrm{arcmin}$, so that the two probe
comparable angular scales and the curvelet advantage reflects directionality
rather than scale selection. \textbf{Black (NRE, expected $95\%$ upper limit):} the neural-ratio sensitivity of Section \ref{sec:NreConst}, i.e.\ the upper limit on $\eta$ conditional on a fixed redshift $z$, shown as a function of $z$. For each of $8000$ wall-free skies the limit is the $\eta$ at which the one-sided statistic $q$ reaches $t_{\rm crit} = 2.71$; the curve is the ensemble mean and the purple band its standard deviation. It lies lowest among our methods because the NRE conditions on the full two-dimensional temperature field and pools the coherent deflection signature across every pixel the wall spans, rather than collapsing the map to a single moment or comparing it to a single resolution element, and it tracks the instrument-noise reference closely.}
    \label{fig:Constraints} 
\end{figure*}

\subsection{SPT Instrumentation Reference}

One marker of detectability is that the wall's lensing distortion is larger than the scale of the instrument resolution. This can mean (1) whether the two deflection peaks flanking the wall can be resolved; or (2) whether the (maximum) deflection amplitude itself exceeds the resolution scale, i.e., $\alpha_\text{max}>\theta_\text{beam}$. Given that we know the distribution of a non-distorted sky, the second interpretation should be a better estimate for the maximum achievable sensitivity. This estimate however is not a hard limit, since coherent sub-pixel distortion over many pixels can still move a primordial sky out of its prior distribution, and thus be detectable at least in principle. Instead, it should indicate the location beyond which sensitivity of any methods rapidly falls. We therefore do not plot this criterion in Figure \ref{fig:Constraints}.

The above argument ignores instrument noise completely, which we now consider. A deflection larger than the pixel size is not observable if the shift in temperature it induces $\delta T = \boldsymbol{\alpha}\cdot\nabla T_{\rm CMB}$ is much smaller than the instrument noise. In Fig.~\ref{fig:Constraints}, we show a curve (blue) set by equating the averaged temperature shift to the mean noise level $\alpha_{\text{min,noise}}\equiv\sigma_\text{noise}/\langle | \nabla T_\text{CMB}|^2\rangle^{1/2}$, where $\sigma_{\rm noise}$ is the
SPT-3G white-noise level per beam at $150\,\mathrm{GHz}$ and $\langle|\nabla T_{\rm CMB}|^2\rangle^{1/2}$ is the rms gradient of the primary sky.
We stress that this is a per-resolution-element reference rather than a sensitivity floor, since methods that probe coherent patterns over many pixels are not necessarily bounded by this curve.

\subsection{Kurtosis sensitivity}
\label{sec:KurtosisConst}
The instrumental boundaries above ask only whether a single number, a separation or an amplitude, crosses an instrumental scale. A statistical detection, by contrast, exploits the full distribution of the lensed field. As described in Section \ref{sec:Curvelet}, the signed curvelet coefficients of a wall-lensed map develop heavy, non-Gaussian tails relative to the approximately Gaussian primordial sky, because the coherent line-like deflection deposits power into a narrow range of curvelet scales and orientations. The excess kurtosis of the signed curvelet coefficients is therefore a natural detection statistic.

To forecast the statistical reach, we generate an ensemble of Gaussian reference skies and a matched ensemble of wall-lensed skies sharing the same primordial realizations, and compare their curvelet-kurtosis distributions with the discriminability statistic \cite{GreenSwets:1966}
\begin{eqnarray}
  d' &=& \max_{b \in \mathcal{B}}\;
  \frac{\big\langle \mathcal{K}_b(T^{\mathrm{obs}}_i)
        -\mathcal{K}_b(T^{\mathrm{prim}}_i)\big\rangle_i}
       {\sigma_{\mathrm{null},\,b}};
  \\
  \sigma_{\mathrm{null},\,b} &\equiv&
  \mathrm{std}_i\!\big[\mathcal{K}_b(T^{\mathrm{prim}}_i)\big] \nonumber
  \label{eq:dprime}
\end{eqnarray}
where $\mathcal{K}_b(T_i^{\rm obs})$ and $\mathcal{K}_b(T_i^{\rm prim})$ are the
kurtoses in channel $b$ of the lensed and null sky computed from the shared
primordial realization $i$, and $\mathcal{B}$ is the channel set: the four
wavelet decomposition levels, or the two curvelet radial bands. The numerator is
a paired difference and the denominator an unpaired scatter, so $d'$ answers the
question of by how many null standard deviations the wall displaces the kurtosis
of a single patch.

We estimate each boundary from an ensemble of $100$ paired
$10^\circ \times 10^\circ$ realizations: each wall-lensed sky is compared
against the Gaussian reference built from the same primordial realization, so
that the pairing cancels cosmic variance and isolates the wall contribution.
The scan runs over a grid of $20$ tensions, $\eta \in [5,100]\,\mathrm{MeV}$ in
steps of $5\,\mathrm{MeV}$, and $120$ redshifts uniformly spaced on
$z \in [1,120]$. At each grid point we build the efficiency-folded deflection
field on the $512^2$ lensing grid, remap all $100$ primordial skies, add an
independent noise realization to each, and evaluate $\mathcal{K}_b$ in all six
channels. The null ensemble is generated once at the start of the scan and held
fixed, so that $\sigma_{\mathrm{null},b}$ is a single per-channel number
throughout and the boundary is not modulated by null-ensemble scatter.

The discriminability $d'$ of (\ref{eq:dprime}) is normalized by the
across-realization scatter of the reference kurtosis, not by
$\sigma_{\rm null}/\sqrt{N}$, so it measures the significance attainable from a
\emph{single} SPT-sized patch, and the $100$ realizations serve only to estimate
the boundary precisely, as $d'$ converges to a fixed value rather than inflating
as $\sqrt{N}$. This makes $d'$ an effect size rather than a $t$-statistic. This
single-patch interpretation is the relevant one for an SPT survey, whose
footprint tiles into of order $20$ such patches.

For each tension $\eta$ we ascend in redshift and record the \emph{highest} $z$
at which $d'$ exceeds threshold, terminating the climb after three consecutive
sub-threshold redshifts to guard against isolated fluctuations near the
boundary. Tensions that yield no detection within the first few redshift steps
are skipped. The search for each $\eta$ resumes from the threshold redshift
found at the previous, lower tension, which imposes by construction a
monotonically non-decreasing frontier $z_\mathrm{det}(\eta)$. This is a
deliberate regularization rather than a result: a stronger wall is detectable to
at least as high a redshift as a weaker one, so enforcing monotonicity
suppresses grid-level noise in the boundary without biasing it.

In accordance with a Bonferroni correction for the look-elsewhere effect
incurred by maximizing over channels, we adopt a one-sided per-channel level
$\alpha_1 = 0.05/M$ for a family-wise level of $5\%$, giving $d' > 1.96$ for the
$M=2$ curvelet bands and $d' > 2.24$ for the $M=4$ wavelet
levels.\footnote{The curvelet threshold coincides numerically with the familiar
two-sided $95\%$ normal quantile; this is a coincidence of $M=2$ and not the
uncorrected value. The corresponding \v{S}id\'{a}k thresholds, obtained from
$\alpha_1 = 1-(1-0.05)^{1/M}$, are $1.95$ and $2.23$, and shift the boundaries
by less than one grid step in $z$.} The larger wavelet threshold reflects the
larger channel set over which the maximum is taken. The locus of these
thresholds defines the curvelet and wavelet detection boundaries.

The channel that attains the maximum is informative in its own right. Among the
curvelet bands the finer $B_2$ band, spanning $2.6$--$4.7\,\mathrm{arcmin}$,
dominates over essentially the whole grid, consistent with the wall depositing
its power at the scale of the void boundary rather than at the scale of the
patch. Among the wavelet levels, level $2$ ($4.7$--$9.4\,\mathrm{arcmin}$)
dominates: level $1$, although nominally matched to the same scales as $B_2$,
is swamped by the white noise, whose flat spectrum places most of its variance
in the finest wavelet sub-bands, while levels $3$ and $4$ are too coarse to
resolve the gradient discontinuity. That the best curvelet band and the best
wavelet level probe comparable angular scales, and that the curvelet nonetheless
reaches deeper, isolates directionality rather than scale selection as the
source of its advantage. This is a direct consequence of the anisotropic,
edge-matching response of the curvelet elements in the context of the line-like
distortion and demagnification along the domain wall plane.

Two simplifications enter these maps. First, we do not convolve with an
instrumental beam; the $1.17\,\mathrm{arcmin}$ pixelization is comparable to the
SPT-3G $150\,\mathrm{GHz}$ beam, so the smallest resolved scale is roughly
correct, but the finest wavelet level in particular retains power that a true
beam would suppress. Since the boundary is set by level $2$ and by the curvelet
$B_2$ band rather than by level $1$, we expect the effect on the quoted curves
to be modest. Second, the noise is taken to be white and homogeneous at
$10\,\mu\mathrm{K}\,\mathrm{arcmin}$, with no atmospheric $1/f$ component, no
inhomogeneous coverage and no foregrounds. Both simplifications push the
forecast in the optimistic direction.

We emphasize that this is a statement of statistical significance, not of
detection power: above the curve, the curvelet kurtosis of an SPT-like ensemble
distinguishes wall-lensed from Gaussian skies with a false-positive probability
below $5\%$. This marks where the method becomes sensitive to such a wall at
this significance, and is an idealized reach rather than an exclusion of real
walls. The kurtosis constraint for curvelets (green) and wavelets (orange) can
be found in Figure \ref{fig:Constraints}.

\subsection{NRE sensitivity}
\label{sec:NreConst}
While the instrumentation and statistical boundaries above compress the map to a single quantity, the NRE instead conditions a network on the full two-dimensional temperature field by use of the learned log ratio $\ln r(x,\Theta)$, and thus exploits the full morphology of the map. Also, while the curvelet and wavelet boundaries are detection thresholds, the NRE directly yields a proper limit. That is, given an observed map of the CMB, we obtain a likelihood-ratio test statistic resulting in a frequentist upper limit on $\eta$ at each scattering redshift $z$. 

Although the kurtosis boundaries are phrased as detection thresholds, they measure essentially the same quantity. For a statistic with a Gaussian null distribution, the median expected one-sided $95\%$ upper limit lies where the wall shifts the statistic by $1.64\,\sigma_{\rm null}$, i.e.\ at $d' = 1.64$, against the $d' = 1.96$ and $2.24$ used for the curvelet and wavelet boundaries, which additionally correct for the choice of channel. Recasting those boundaries as upper limits would therefore lower them only modestly and would not change their ordering relative to the NRE.

\begin{figure}
    \centering
    \includegraphics[width=\linewidth]{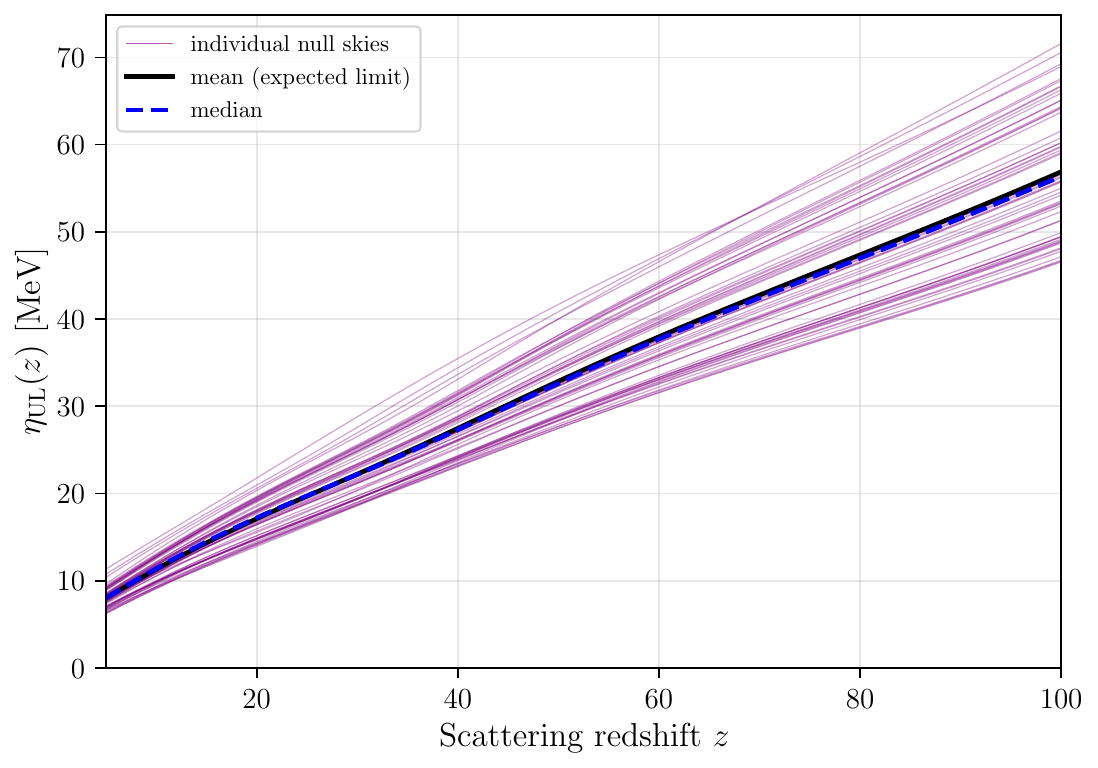}
    \caption{Conditional upper limits $\eta_{\rm UL}(z)$ for $60$ individual null skies (thin purple), with the ensemble mean (black), the expected limit of Figure \ref{fig:Constraints}, and the median (dashed blue).}
    \label{fig:NRENull}
\end{figure}

The sensitivity is computed from $8000$ maps without a wall, simulated through the same forward model as the training set. For each null map we evaluate $\ln r(x;z,\eta)$ on a grid of $61$ redshifts $z \in [5,100]$ and $400$ values $\eta \in [0,100]\,\mathrm{MeV}$, and at each redshift record the $\eta$ above $\hat\eta$ at which the one-sided statistic $q$ of (\ref{eq:qstat}) reaches $t_{\rm crit} = 2.71$, i.e.\ where $\ln r$ has fallen $1.35$ below its maximum at that redshift. On a null sky the maximum lies on the low-$\eta$ plateau where $\ln r$ is flat, at $\hat\eta \lesssim 15\,\mathrm{MeV}$ in the median, and the crossing is insensitive to its exact position. Tracing the crossing across $z$ gives, for each sky, a curve $\eta_{\rm UL}(z)$ (Figures \ref{fig:NREInject} and \ref{fig:NRENull}): the tensions that sky excludes at $95\%$ confidence for a wall at redshift $z$. Each point is a limit on $\eta$ conditional on the redshift, with $z$ labelling the hypothesis rather than being estimated, which is the standard construction for an exclusion curve. The mean of $\eta_{\rm UL}(z)$ over the ensemble is the expected limit, shown as the black curve in Figure \ref{fig:Constraints}, and its standard deviation is the purple band; the limit ranges from $8.0 \pm 1.0\,\mathrm{MeV}$ at $z = 5$ to $56.9 \pm 6.4\,\mathrm{MeV}$ at $z = 100$, and the median and $16$--$84\%$ range agree with these to within $1\%$ and $10\%$. Since the ensemble contains no walls, the band reflects how the limit attainable from a single SPT-sized patch varies with the primordial and noise realization, and not the range of $\eta$ that would be recovered for a wall that is present.

At fixed $z \gtrsim 20$ the NRE limit lies a factor $1.3$--$1.9$ below the kurtosis boundaries in $\eta$. The gain over the kurtosis boundaries is considerably larger than the difference between the wavelet and curvelet boundaries, which reflects directionality alone, so the full lensing morphology carries most of the information. Against the noise reference, however, the NRE does not reach deeper: the two curves nearly coincide across the redshift range, the NRE limit lying within $\sim 5\%$ of the noise criterion and marginally above it at the highest redshifts. The NRE therefore attains calibrated sensitivity to essentially any wall whose coherent imprint rises above the SPT-3G-like noise floor we adopt.

Although the statistical wavelet and curvelet methods also utilize the full temperature field, their reach is ultimately weaker than that of the NRE. The kurtosis, though evaluated on the full map, compresses it to a single fixed moment and retains only the heavy tail weight of the coefficient distribution, discarding the spatial coherence of the deflection that most distinguishes a wall. The NRE instead keeps the full field and pools that coherent signal across every pixel the wall spans, which is what carries its forecast deeper.

\section{\label{sec:Conclusion}Conclusion}

We have investigated the imprint of dark, gravitationally interacting domain
walls on the cosmic microwave background, from the underlying deflection
physics through to forecasted observational constraints. Starting from the
relativistic repulsion of matter by a planar wall, we derived the surrounding
density profile, an evacuated void bounded by an overdense shell, and computed
the resulting photon deflection field analytically for both aligned and
arbitrarily inclined walls. A central qualitative result is that the deflection
produces coherent, line-like, strongly non-Gaussian distortions in the
temperature field, with a characteristic two-peak structure set by the
relativistic displacement scale $\psi(z)$ and a sharp dependence on the wall
inclination that vanishes entirely for the face-on configuration.

Because these signatures are localised and anisotropic, they are poorly
captured by two-point statistics and instead call for methods sensitive to
coherent structure. We pursued two complementary approaches. The first is a
curvelet-based analysis, whose directional multiscale basis is naturally
matched to the wall's line-like geometry and whose signed-coefficient kurtosis
distinguishes lensed from Gaussian skies. The second is simulation-based
inference, in which neural networks learn directly from forward simulations.
Within the latter we trained both a neural posterior estimator, which recovers
the joint posterior over $z$, $\eta$, and $\theta$ and exposes the
degeneracy between $z$ and $\eta$ from a single patch, while recovering $\theta$
independently, and a neural ratio estimator, which provides a likelihood-ratio test statistic, with an asymptotic threshold verified on simulations, suited to limit-setting.

Casting these methods as forecasts on an SPT-like survey, we mapped the region
of the $(z,\eta)$ plane in which domain walls become detectable by each method.
The curvelet and wavelet statistical thresholds populate a common
$\eta \sim 10$ to $100\,\mathrm{MeV}$ band over the accessible redshift range,
while the neural-ratio sensitivity curve, which exploits the full map
morphology, reaches a factor $1.3$--$1.9$ lower in $\eta$ for $z \gtrsim 20$. That curve in turn tracks the instrumental
noise reference closely across the whole redshift range. The convergence of
these independently constructed curves is a useful internal consistency check,
and it identifies the detectability of these walls as limited by statistical
power for the moment-based methods and by the noise floor for the best of them,
rather than by angular resolution. The neural-ratio curve lies notably deeper
in the parameter space than the single-statistic forecasts, reflecting that
exploiting the coherent structure of the lensing signal yields the strongest
sensitivity.

Several limitations qualify these forecasts and motivate future work, and they
are the reason the curves should be read as idealized sensitivities rather than
exclusions. The maps used throughout are idealized: a single, optimally aligned
wall is imprinted on an otherwise Gaussian sky, with no $\Lambda$CDM lensing,
foregrounds, point sources, inhomogeneous noise or beam systematics. The curves
therefore represent the reach of each method under ideal conditions, and folding
realistic instrumental and astrophysical effects into the maps would move them
toward weaker sensitivity. This is a natural next step. The single-patch
inference is fundamentally limited by the degeneracy between $z$ and $\eta$,
which a multi-patch or full-sky analysis could partially break. Our treatment
also assumes the scaling expectation of approximately one wall per Hubble
volume, and neglects the integrated Sachs--Wolfe contribution expected to
dominate for near-face-on walls; a complete treatment of the time-dependent
gravitational potential, together with the redshift-dependent overclosure
constraint that bounds the viable parameter space from above, would sharpen the
cosmological picture. The evacuated void and overdense shell should likewise imprint a
distinctive signature on the redshifted 21 cm brightness temperature, in analogy with the
21 cm signature of cosmic string wakes \cite{Brandenberger:2010,Hernandez:2011,Hernandez:2012},
which would provide a probe complementary to the lensing analysis presented here. Finally, the curves presented here are idealized
sensitivity forecasts, and demonstrate as a proof of concept that CMB lensing
can be used to find domain walls. Applying these pipelines to real CMB data from
current and next-generation experiments such as SPT-3G \cite{SPT3G:2021}, the Simons Observatory \cite{SimonsObs:2019},
CMB-S4 \cite{CMB-S4:2016}, and LiteBIRD \cite{LiteBIRD:2022} remains the essential next step, and would convert the
detectability map derived here into genuine bounds on, or a detection of,
cosmological domain walls.

\begin{acknowledgments}
The research at McGill University is supported in part by funds from Natural Sciences and Engineering Research Council of Canada (NSERC) and from the Canada Research Chair Program. BG was supported by a Master’s and Doctoral Research Scholarship from Fonds de recherche du Québec. BG is deeply grateful to Katelin Schutz and the DMcGill group for their mentorship and support throughout this work.
YS acknowledges support from a Trottier Space Institute Fellowship at McGill and support from a NSERC Subatomic Physics Discovery Grant.
Lastly, we thank Keshav Dasgupta for useful comments on this article.

\end{acknowledgments}

\clearpage
\appendix

\onecolumngrid

\section{\label{sec:AppDeflection}Geometric Kernels of the Deflection Angle}
We consider a domain wall of finite extent $L \times L$ rotated by an angle $\theta$ about the $y$-axis. A rotation about $z$ merely rotates the observational window and produces no new physics, so $\theta$ fully parameterizes the arbitrary orientation of a domain wall.

A CMB photon crossing the matter-wake created by the domain wall is deflected as a result of the contrast to the homogeneous background, $\Sigma \equiv \rho - \bar{\rho}$. In the weak-field limit, with the Born approximation, the photon is taken to travel in a straight line through the matter profile and so the impulse transverse to the photons path can be integrated to obtain the deflection angle
\begin{equation}
  \vec \alpha(x_0,y_0)=\frac{4G}{c^{2}}\!\int\!dx'\,dy'\,dz'\;\Sigma(x',z')\,
  \frac{(x_0-x',y_0-y')}{(x_0-x')^{2}+(y_0-y')^{2}},
  \label{eq:born}
\end{equation}
Introduce coordinates aligned with the wall normal,
\begin{equation}
  u = x'\cos\theta-z'\sin\theta,\qquad w = x'\sin\theta+z'\cos\theta,
\end{equation}
with inverse $x'=u\cos\theta+w\sin\theta$, $z'=-u\sin\theta+w\cos\theta$. The rotation has
unit Jacobian, $dx'\,dz'=du\,dw$. By construction the matter distribution depends only on
the perpendicular distance to the wall, $\Sigma(x',z')=\Sigma(u)$, while the finite
wall extent restricts the in-plane coordinates,
\begin{equation}
  |w|<\tfrac{L}{2},\qquad |y'|<\tfrac{L}{2},\qquad u\in(-r_H,\,r_H).
\end{equation}
The contrast is the frozen relativistic wake profile (the evacuated void carved out by the
wall's repulsion, bounded by the displaced-matter overdensity). It is even in $u$ and, with
$\psi$ the comoving half-width of the void and $\bar\rho$ the background density at the
freeze redshift,
\begin{equation}
  \Sigma(u)=
  \begin{cases}
    -\,\bar\rho, & |u|\le\psi,\\[2pt]
    \rho(u)-\bar\rho, & \psi<|u|<r_H,\\[2pt]
    0, & |u|\ge r_H,
  \end{cases}
  \label{eq:contrast}
\end{equation}
where on the shell $\rho(u)$ is the numerically evolved profile, decaying to $\bar\rho$ at
the horizon so that $\Sigma\to0$ there. Mass conservation guarantees
$\int_{-r_H}^{r_H}\Sigma(u)\,du=0$. Nothing in what follows assumes a closed form for
$\rho(u)$: the deflection construction is agnostic to how the profile was obtained, which is
precisely what lets the relativistic-matter wake reuse the kernels derived for the static
profile.

Substituting the rotation into (\ref{eq:born}) and abbreviating
$s\equiv x_0-u\cos\theta-w\sin\theta$, the $y'$ integral is elementary,
\begin{equation}
  \int_{-L/2}^{L/2}\!dy'\;\frac{s}{s^{2}+(y_0-y')^{2}}
  =\arctan\frac{Y_+}{s}+\arctan\frac{Y_-}{s},
  \qquad Y_\pm\equiv\tfrac{L}{2}\pm y_0 .
\end{equation}
The remaining $w$ integral is converted to an $s$ integral via $ds=-\sin\theta\,dw$, with
$w=\mp L/2$ mapping to $s=s_\pm$ where
\begin{equation}
  s_\pm\equiv x_0-u\cos\theta\pm\tfrac{L}{2}\sin\theta .
\end{equation}
Using $\int\!\bigl[\arctan\tfrac{Y_+}{s}+\arctan\tfrac{Y_-}{s}\bigr]ds
= s\bigl(\arctan\tfrac{Y_+}{s}+\arctan\tfrac{Y_-}{s}\bigr)
+\tfrac{Y_+}{2}\ln(s^{2}+Y_+^{2})+\tfrac{Y_-}{2}\ln(s^{2}+Y_-^{2})\equiv g(s)$, the two
in-plane integrals collapse to a single closed-form kernel,
\begin{equation}
  \alpha_x(x_0,y_0)=\frac{4G}{c^{2}}\!\int_{-r_H}^{r_H}\!du\;\Sigma(u)\,K_x(u),
  \qquad
  K_x(u)=\frac{1}{\sin\theta}\bigl[g(s_+)-g(s_-)\bigr].
  \label{eq:Kx}
\end{equation}
The crucial point is that the $y'$ and $w$ integrals acted only on the geometric kernel and
never on $\Sigma(u)$: $K_x(u)$ depends solely on $(x_0,y_0,u,\theta,L)$ and is therefore
\emph{the same for any radial profile}. Repeating the calculation for the $y$ component, the
$y'$ integral now yields $\tfrac12\ln\!\bigl[(s^{2}+Y_+^{2})/(s^{2}+Y_-^{2})\bigr]$ and the
$w$ integral gives
\begin{align}
  \alpha_y(x_0,y_0)&=\frac{4G}{c^{2}}\!\int_{-r_H}^{r_H}\!du\;\Sigma(u)\,K_y(u),\nonumber\\
  K_y(u)&=\frac{1}{\sin\theta}\bigl[q(s_+)-q(s_-)\bigr],\qquad
  q(s)=\frac{s}{2}\ln\frac{s^{2}+Y_+^{2}}{s^{2}+Y_-^{2}}
       +Y_+\arctan\frac{s}{Y_+}-Y_-\arctan\frac{s}{Y_-}.
  \label{eq:Ky}
\end{align}
Finally, since $\Sigma$ is even in $u$ while the kernels are not, the integral over the
two shells folds onto the positive half-line,
\begin{equation}
  \boxed{\;\alpha_{x,y}(x_0,y_0)=\frac{4G}{c^{2}}\int_{0}^{r_H}\!du\;\Sigma(u)\,
  \bigl[K_{x,y}(u)+K_{x,y}(-u)\bigr].\;}
  \label{eq:final}
\end{equation}
(\ref{eq:final}) is exact and holds for any even profile. The relativistic FRLW wake is
known only as a numerical profile and we evaluate (\ref{eq:final}) directly. We split the range at the void edge,
\begin{equation}
  \int_0^{r_H}\!du\;\Sigma(u)\,[\cdots]
  =\underbrace{(-\bar\rho)\!\int_0^{\psi}\!du\,[\cdots]}_{\text{void, constant}}
   +\underbrace{\int_{\psi}^{r_H}\!du\,\bigl(\rho(u)-\bar\rho\bigr)[\cdots]}_{\text{shell, numerical}},
\end{equation}
so that the jump in $\Sigma$ at $u=\psi$ falls on a panel boundary, and apply
Gauss--Legendre quadrature to each interval, with the $(x_0,y_0)$ grid carried vectorially so
that the per-node cost is one grid evaluation of $K_{x,y}(u)+K_{x,y}(-u)$. The kernels are
evaluated with the principal branch of $\arctan$, which is the continuous antiderivative used
in deriving $g$ and $q$; the four-quadrant $\operatorname{atan2}$ must \emph{not} be
substituted, as its $2\pi$ branch jumps reintroduce spurious discontinuities at $s=0$.

The deflection of (\ref{eq:final}) is the lens-frame angle. The deflection imprinted on
the observed CMB is reduced by the lensing efficiency,
\begin{equation}
  \vec{\alpha}_{\rm obs}=\frac{D_{\ell s}}{D_s}\,\vec{\alpha},
  \qquad \frac{D_{\ell s}}{D_s}=1-\frac{\chi_\ell}{\chi_s},
\end{equation}
with $\chi_\ell,\chi_s$ the comoving distances to the wall and to the last-scattering surface,
and is applied to the temperature field as a remapping
$T_{\rm obs}(\vec{\theta})=T_{\rm prim}\!\bigl(\vec{\theta}
-\vec{\alpha}_{\rm obs}\bigr)$ after mapping the comoving deflection to angular
coordinates through the angular-diameter distance, $(x_0,y_0)=D_A\, (\theta_x, \theta_y)$,
$D_A=a(z)\chi(z)$.

Two limits validate the construction. For the face-on wall, $\theta\to\pi/2$, the line of
sight runs along $u$ and the projected contrast is the total column,
$\Sigma=\int_{-r_H}^{r_H}\Sigma(u)\,du=0$ by mass conservation, the
deflection therefore vanishes identically, and the wall is invisible to lensing (any residual
signal is of integrated Sachs--Wolfe type and is neglected here). For the aligned wall,
$\theta\to0$, $s_\pm\to x_0\mp\tfrac{L}{2}\sin\theta\to x_0$ along the projection and
(\ref{eq:final}) reduces to the edge-on $L\times L$ sheet, recovering the
one-dimensional result by completing the $z'$ integral trivially. Between these, the inclination enters only through $s_\pm$ and the
overall $1/\sin\theta$, so a single quadrature spans all orientations.

\newpage

\section{\label{sec:AppCalibration}Posterior Calibration of the Neural Posterior Estimator}

\begin{figure}[h]
    \centering
    \includegraphics[width=0.5\linewidth]{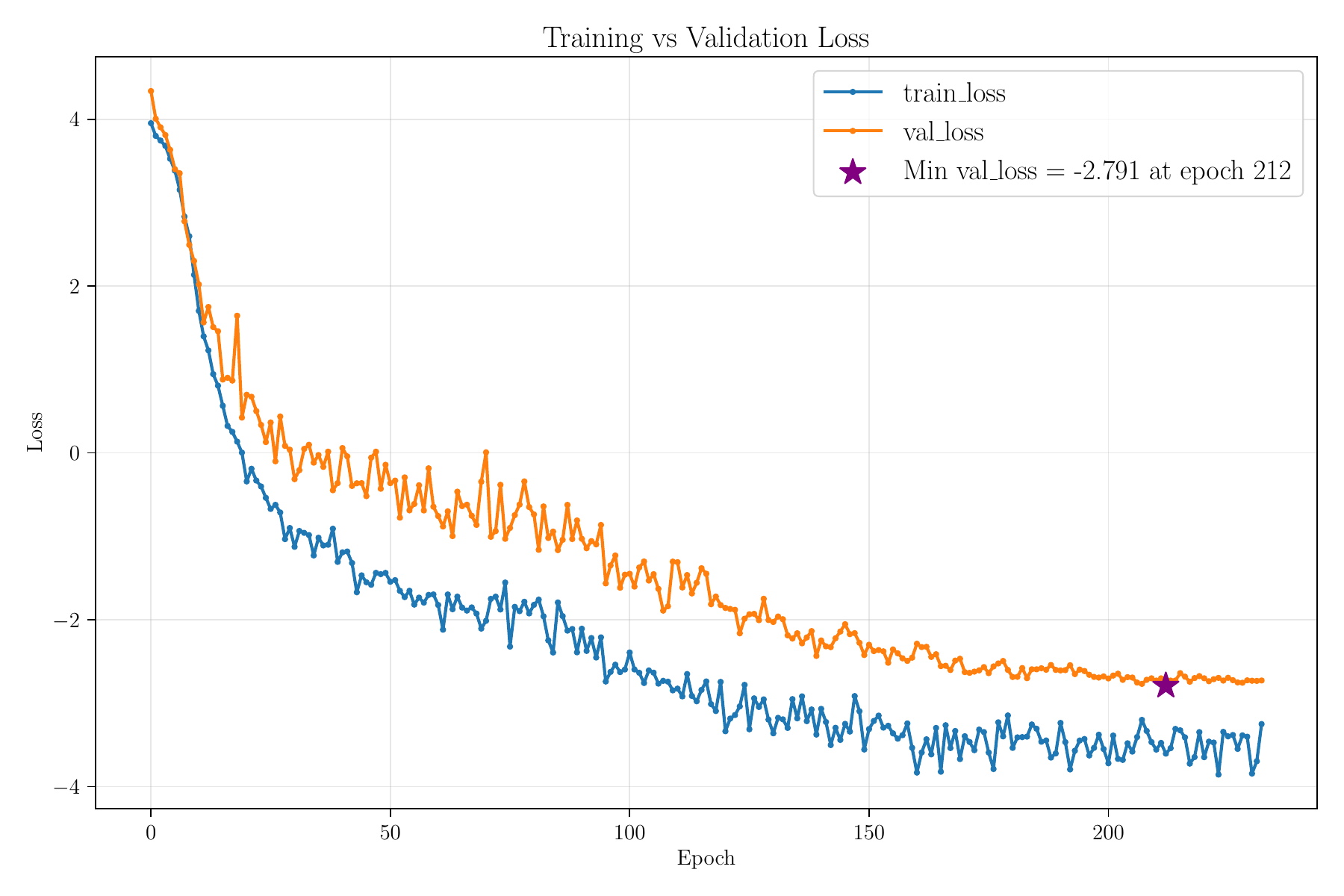}
    \caption{Training and validation loss for the NPE (negative log-likelihood of
(\ref{eq:npe-loss})) against epoch. The training loss is evaluated on
the maps used to fit the network and the validation loss on held-out maps.
Convergence of the two curves indicates the network has been trained close to
the limit set by the available data; the epoch of minimum validation loss
defines the checkpoint used throughout.}
\label{fig:NPELoss}
\end{figure}

\begin{figure}[h]
    \centering
\includegraphics[width=0.5\linewidth]{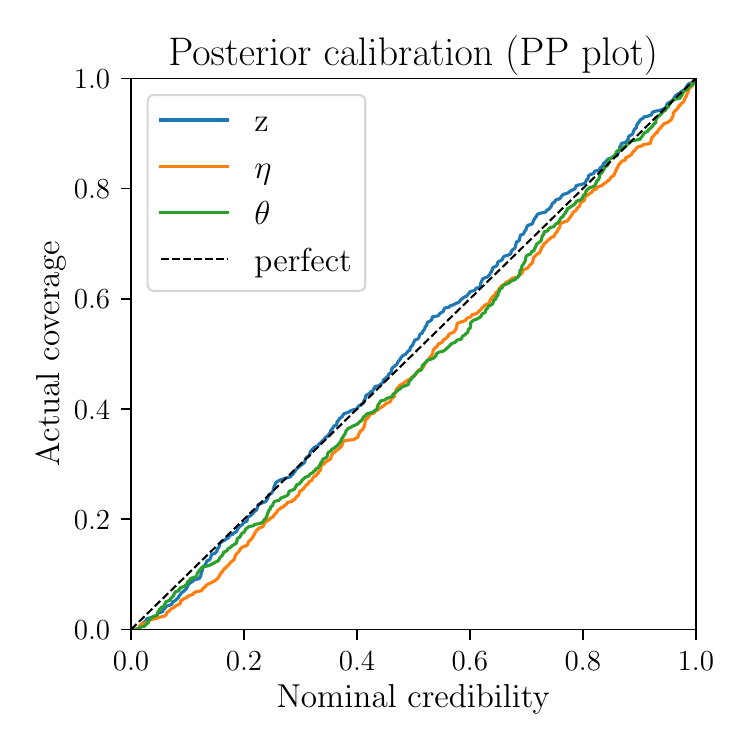}
    \caption{
    Posterior calibration of the trained NPE, shown as a
probability--probability plot. For each parameter, the empirical coverage --- the
fraction of the $200$ validation maps whose true value falls inside the central
credible interval of the stated nominal credibility --- is plotted against that
nominal credibility. The dashed diagonal is perfect calibration; curves above it
over-cover (conservative intervals) and curves below it under-cover
(over-confident intervals). All three parameters track the diagonal to within a
few percent, with $z$ marginally conservative and $\eta$ and $\theta$
marginally over-confident. For $N=200$ maps the binomial uncertainty on the
coverage is $\sqrt{X(1-X)/N} \simeq 3.5\%$ near mid-range, comparable to the
observed departures.}
    \label{fig:PPplot}
\end{figure}

\newpage

\section{Calibration of the Neural Ratio Estimator}
\label{sec:AppNRECalibration}

We report here a check of the asymptotic thresholds
used in Sections \ref{sec:NRE} and \ref{sec:NreConst}. For each of the $4096$
validation maps,\footnote{These maps were never used to update the network
weights, but they were used to select the checkpoint and to schedule the
learning rate. $377$ of them are null skies.} we hold $z$ at its true value and
evaluate $\ln r(x;z,\eta)$ on a grid of $400$ values of
$\eta \in [0,100]\,\mathrm{MeV}$. From each curve we record the test statistic
at the truth, $t = 2[\ln r_{\max} - \ln r(x;z,\eta_{\rm true})]$, and the
credibility of the smallest highest-density region, under a uniform prior on
$\eta$, that contains $\eta_{\rm true}$.

\begin{figure*}
    \centering
    \includegraphics[width=\linewidth]{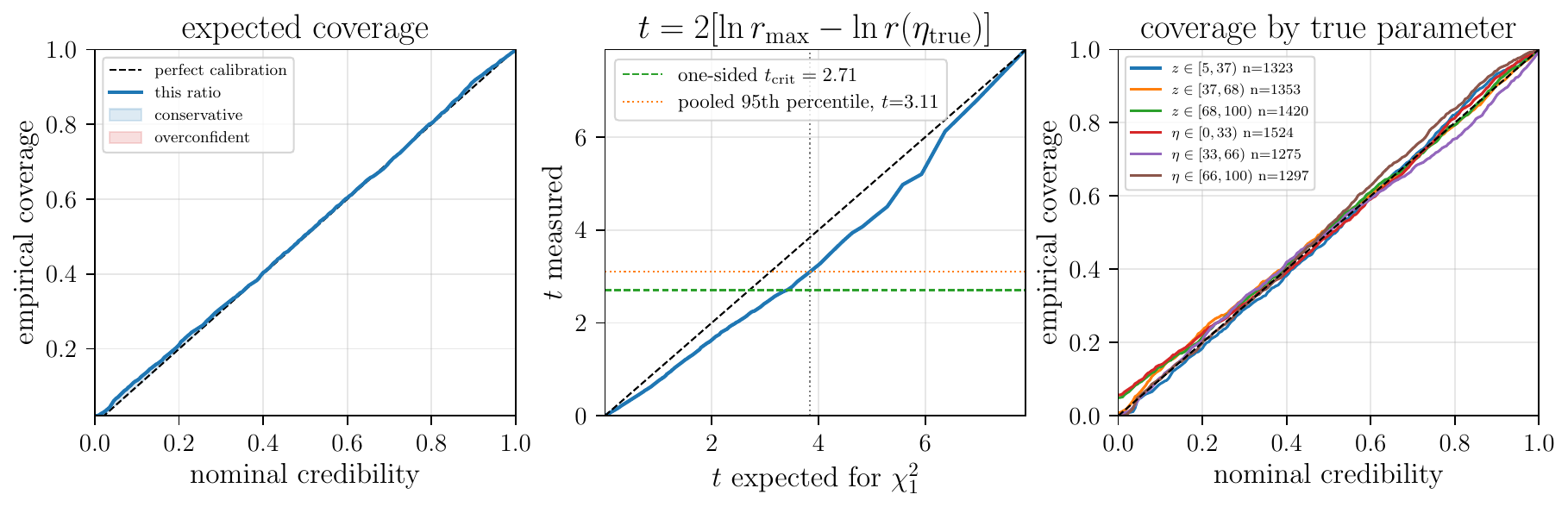}
    \caption{Calibration of the neural ratio estimator on $4096$ validation
maps, with $z$ held at its true value. \emph{Left:} expected coverage of the
highest-density $\eta$ regions under a uniform prior on $\eta$, against their
nominal coverage, and the dashed diagonal is perfect calibration. \emph{Centre:} measured $t = 2[\ln r_{\max} - \ln r(\eta_{\rm true})]$ against the value expected for $\chi^2_1$ at the same percentile, from the 1st to the 99.5th; points below the dashed diagonal mean $t$ is smaller than $\chi^2_1$ predicts. \emph{Right:} expected coverage split by true $z$ and
true $\eta$. The non-zero intercepts of two curves come from null skies whose
$\ln r$ peaks exactly at $\eta_{\rm true} = 0$. }
    \label{fig:NRECalib}
\end{figure*}

The result is shown in Figure \ref{fig:NRECalib}. The expected coverage (left
panel) follows the diagonal closely, with $97.7\%$ coverage at a nominal
$97.5\%$, so that on average the ratio is neither overconfident nor
conservative. Pooled over the whole prior, the $95$th percentile of $t$ (centre panel) is $3.11 \pm 0.11$, where the error is the standard deviation over $500$ bootstrap resamplings, below the $3.84$ of $\chi^2_1$.

Binning by the true parameters (right panel and Table \ref{tab:NRECalib})
shows that this pooled value is a mixture. The interior band
$\eta \in [33,66)\,\mathrm{MeV}$ gives $\Delta\ln r = 2.00 \pm 0.14$, consistent
with the Wilks value $1.92$, and is mildly overconfident, with $95.8\%$
coverage at a nominal $97.5\%$. The two outer bands give $\Delta\ln r \approx
1.35$, consistent with the value expected where the truth lies on a boundary
of the parameter space, for which $t$ follows
$\tfrac{1}{2}\chi^2_0 + \tfrac{1}{2}\chi^2_1$ with $95$th percentile $2.71$
\cite{Chernoff:1954}. In the lowest band, null skies and walls below the
detection scale, for which $\ln r$ is nearly flat in $\eta$, further contribute
small values of $t$. The dependence on $z$ is weaker and non-monotonic. A single pooled percentile therefore has no clean interpretation, and we use the asymptotic one-sided threshold of Section \ref{sec:NRE} instead.

\begin{table}[h]
\caption{Half the $95$th percentile of $t$, $\Delta\ln r = t_{95}/2$, and expected coverage at
a nominal $97.5\%$, binned by the true parameters of the validation maps.}
\label{tab:NRECalib}
\begin{ruledtabular}
\begin{tabular}{lccc}
Bin & $n$ & $\Delta\ln r$ & Coverage \\
\hline
$z \in [5,37)$         & 1323 & $1.37 \pm 0.09$ & 0.982 \\
$z \in [37,68)$        & 1353 & $1.76 \pm 0.11$ & 0.973 \\
$z \in [68,100)$       & 1420 & $1.53 \pm 0.08$ & 0.978 \\
$\eta \in [0,33)$ MeV  & 1524 & $1.37 \pm 0.06$ & 0.982 \\
$\eta \in [33,66)$ MeV & 1275 & $2.00 \pm 0.14$ & 0.958 \\
$\eta \in [66,100)$ MeV& 1297 & $1.33 \pm 0.08$ & 0.991 \\
\hline
All                    & 4096 & $1.56 \pm 0.05$ & 0.977 \\
\end{tabular}
\end{ruledtabular}
\end{table}

For the upper limit, only failures with $\eta_{\rm true} > \hat\eta$ count, and the relevant statistic is the one-sided $q$ of Section \ref{sec:NRE}. On the $3719$ validation maps containing a wall, the upper limit at $t_{\rm crit} = 2.71$ covers the true $\eta$ in $97.3\%$ of cases, against a nominal $95\%$. Binned in the true tension, the coverage is $95.5\%$ in the interior band $\eta \in [33,66)\,\mathrm{MeV}$, consistent with the nominal value within the binomial error of $0.6\%$, and $98.9\%$ and $97.6\%$ in the lower and upper bands, where the prior boundaries and the flat low-$\eta$ plateau make the limit conservative; binned in redshift it lies between $96.4\%$ and $97.8\%$. The choice of threshold has little effect on the sensitivity curve: replacing $\Delta\ln r = 1.35$ by the pooled $1.56$ or by the two-sided $1.92$ raises the expected limit by $2$--$3\%$ and $6$--$7\%$, respectively, at every redshift.

For the joint contour of Figure \ref{fig:NREInject}, the $95$th percentile of $t_{\rm 2D} = 2[\ln r_{\max} - \ln r(x;z_{\rm true},\eta_{\rm true})]$ over $2000$ validation maps on the same $96\times96$ grid is $4.24$, below the $5.99$ of $\chi^2_2$, so the asymptotic contour at $\Delta\ln r = 3.00$ is conservative.

\newpage

\bibliography{References}%

\end{document}